\documentclass[sigconf]{acmart}

\usepackage{float}
\usepackage{textcomp}
\usepackage{tabularx}
\usepackage{tikz}
\usetikzlibrary{arrows.meta, positioning, fit, backgrounds, calc}
\usepackage{pgfplots}
\pgfplotsset{compat=1.18}
\usepackage{algorithm}
\usepackage{algpseudocode}
\newcolumntype{Y}{>{\raggedright\arraybackslash}X}

\setcopyright{none}
\renewcommand\footnotetextcopyrightpermission[1]{}

\newcommand{\baselineB}{Time Headway}

\makeatletter
\NewDocumentEnvironment{IEEEkeywords}{+b}{\gdef\@keywords{#1}}{}
\makeatother

\begin{document}

\title{Drive the Thoughts: Runtime Monitoring of VLA Reasoning-Trajectory Consistency}

\author{Tian Yu}
\affiliation{%
  \institution{University of Virginia}
  \city{Charlottesville}
  \state{Virginia}
  \country{USA}}
\email{buc9hh@virginia.edu}

\author{Lu Feng}
\affiliation{%
  \institution{University of Virginia}
  \city{Charlottesville}
  \state{Virginia}
  \country{USA}}
\email{lu.feng@virginia.edu}

\author{Sebastian Elbaum}
\affiliation{%
  \institution{University of Virginia}
  \city{Charlottesville}
  \state{Virginia}
  \country{USA}}
\email{selbaum@virginia.edu}

\begin{abstract}
\hypersetup{hidelinks}
Autonomous vehicles (AVs) operate in complex environments where failures are consequential. Sophisticated machine learning models for perception and planning are key to overcoming at least part of that complexity, but their black-box nature  complicates validation and verification (V\&V). The recent integration of vision-language-action (VLA) models into AVs introduces a unique opportunity. Besides generating trajectories, these models produce an explicit chain-of-thought (CoT) explaining their underlying decisions. We posit that this CoT provides a candidate specification to cross-check model outputs and detect inconsistencies that may expose unsafe behavior. This paper assesses whether CoTs from a recent open driving VLA can support such checking. We curate DriveAlignBench, a specialized dataset from NVIDIA's Alpamayo 1.5 VLA for AVs containing 150 CoT--trajectory pairs, which we manually annotate for reliability, CoT--trajectory consistency, and trajectory safety. Our analysis reveals that 33.3\% of CoTs are unreliable. Among reliable CoTs, the generated trajectory is inconsistent with the CoT in 26\% of cases. To detect such cases, we propose integrating a CoT--trajectory consistency check into a runtime monitor. The check is nontrivial because CoTs express open-vocabulary, scene-relative driving maneuvers, while trajectories are low-level ego-motion sequences whose semantics depend on road geometry and motion context. To bridge this gap, we develop a family of automated consistency monitors. The strongest, the lane-aware monitor, adds lane-relative evidence to the input of its judge, the model that scores consistency. This raises $F_1$ for all four judges we evaluate, from $0.56$--$0.62$ with raw waypoints to $0.65$--$0.82$, suggesting that evidence representation can improve detection across judges. We also find that, for reliable CoTs, inconsistency is associated with unsafe behavior: 27\% of human-labeled inconsistent trajectories are unsafe, compared to 3\% of consistent ones, and the lane-aware monitor flags most of these unsafe cases. We release DriveAlignBench, the monitor implementations, and per-run results at \url{https://github.com/less-lab-uva/drive-the-thoughts}.
\end{abstract}
 
\begin{IEEEkeywords}
Runtime monitoring, autonomous vehicles, vision-language-action models, chain-of-thought reasoning, reasoning--action consistency, LLM-as-judge, software testing and verification
\end{IEEEkeywords}

\ccsdesc[500]{Software and its engineering~Software verification and validation}
\ccsdesc[300]{Computer systems organization~Robotic autonomy}

\begin{teaserfigure}
\centering
\definecolor{tzInk}{HTML}{23272E}
\definecolor{tzMute}{HTML}{6B7280}
\definecolor{tzRule}{HTML}{C5CAD3}
\definecolor{tzVLA}{HTML}{39414D}
\definecolor{tzJudge}{HTML}{4F55A5}
\definecolor{tzJudgeF}{HTML}{E8E9F7}
\definecolor{tzLane}{HTML}{14806F}
\definecolor{tzPath}{HTML}{E07B1A}
\definecolor{tzOK}{HTML}{1B7837}
\definecolor{tzBad}{HTML}{B2182B}
\begin{tikzpicture}[x=1cm,y=1cm,font=\scriptsize,text=tzInk,line cap=round,
  flow/.style={-{Stealth[length=1.6mm,width=1.3mm]},line width=.55pt,draw=tzInk!80},
  title/.style={font=\footnotesize\bfseries,anchor=base west,inner sep=0},
  frame/.style={rounded corners=2pt,draw=tzRule,line width=.45pt,fill=white},
  stepnum/.style={circle,inner sep=0,minimum size=3.1mm,text=white,font=\scriptsize\bfseries},
  read/.style={inner sep=0,font=\scriptsize\bfseries,text=tzPath!80!black},
  quote/.style={anchor=west,text width=2.78cm,align=left,inner sep=0,font=\footnotesize\itshape,
    execute at begin node={\hyphenpenalty=10000\exhyphenpenalty=10000}}]
\def\rA{-1.85}\def\rB{-3.85}\def\rh{.93}
\def\xS{2.55}\def\xV{2.85}\def\xVe{3.37}\def\xC{3.67}\def\xM{6.95}\def\xMe{10.2}
\def\xE{10.47}\def\xEe{13.07}\def\xJ{13.32}\def\xJe{14.24}\def\xF{14.5}\def\xFe{17.6}
\node[title] at (0,-.62) {Scene};
\node[stepnum,fill=tzVLA] at (\xC+.15,-.56) {1};
\node[title] at (\xC+.37,-.62) {Is the CoT reliable?};
\draw[tzVLA,line width=.6pt] (\xC,-.76) -- (\xM-.1,-.76);
\node[stepnum,fill=tzLane] at (\xM+.15,-.56) {2};
\node[title] at (\xM+.37,-.62) {Does the trajectory realize the CoT?};
\node[title,anchor=base east,text=tzLane!85!black] at (\xJe,-.62) {Lane-aware monitor};
\draw[tzLane,line width=.6pt] (\xM,-.76) -- (\xJe,-.76);
\node[title] at (\xF,-.62) {Monitor verdict};
\foreach \r/\k/\c/\o in {\rA/a/tzOK/-.3,\rB/b/tzBad/-.1}{
  \begin{scope}
    \clip[rounded corners=2pt] (0,\r+\rh) rectangle (\xS,\r-\rh);
    \node[anchor=north west,inner sep=0] at (\o,\r+\rh) {\includegraphics[width=3.05cm,height=1.86cm]{figures/teaser_\k_cam.jpg}};
  \end{scope}
  \draw[\c,line width=.8pt,rounded corners=2pt] (0,\r+\rh) rectangle (\xS,\r-\rh);
  \node[fill=white,rounded corners=1pt,inner sep=1.5pt,font=\scriptsize\bfseries,anchor=north west] at (.07,\r+.86) {(\k)};
  \draw[flow] (\xS,\r) -- (\xV,\r);
  \draw[flow] (\xVe,\r) -- (\xC,\r);
  \node[frame,anchor=north west,minimum width=6.53cm,minimum height=1.86cm,inner sep=0] at (\xC,\r+\rh) {};
  \begin{scope}
    \clip[rounded corners=2pt] (\xM,\r+.92) rectangle (\xMe,\r-.92);
    \node[anchor=north west,inner sep=0] at (\xM,\r+\rh) {\includegraphics[width=3.25cm,height=1.86cm]{figures/teaser_\k_map.pdf}};
  \end{scope}
  \draw[tzRule,line width=.4pt] (\xM,\r+\rh) -- (\xM,\r-\rh);
  \draw[flow] (\xMe+.01,\r) -- (\xE,\r);
  \begin{scope}
    \clip[rounded corners=2pt] (\xE,\r+\rh) rectangle (\xEe,\r-\rh);
    \node[anchor=north west,inner sep=0] at (\xE,\r+\rh) {\includegraphics[width=2.6cm,height=1.86cm]{figures/teaser_\k_lane.pdf}};
  \end{scope}
  \draw[tzLane,line width=.5pt,rounded corners=2pt] (\xE,\r+\rh) rectangle (\xEe,\r-\rh);
  \draw[flow] (\xEe,\r) -- (\xJ,\r);
  \draw[flow] (\xJe,\r) -- (\xF-.01,\r);
}
\draw[tzRule,fill=white,line width=.45pt,rounded corners=1.5pt] (\xV,-.92) rectangle (\xVe,-4.78);
\node[rotate=90,inner sep=0,font=\scriptsize\bfseries] at ({(\xV+\xVe)/2},-2.85) {Driving VLA};
\node[rounded corners=2pt,draw=tzJudge,line width=.45pt,fill=tzJudgeF,text=tzInk,minimum width=.92cm,minimum height=3.86cm,align=center,inner sep=0,font=\footnotesize\bfseries] at ({(\xJ+\xJe)/2},-2.85) {Judge\\[4pt]{\scriptsize\mdseries compares}\\{\scriptsize\mdseries CoT and}\\{\scriptsize\mdseries evidence}};
\node[anchor=west,text width=3.02cm,align=left,inner sep=0] at (\xC+.12,\rA) {\textbf{CoT (a)}\\[1.5pt]{\itshape ``Change lane to the left due to slower traffic in front and a clear gap behind the lead sedan, \ldots''}};
\node[anchor=west,text width=3.02cm,align=left,inner sep=0] at (\xC+.12,\rB) {\textbf{CoT (b)}\\[1.5pt]{\itshape ``Lane change to the left due to slower vehicles ahead in the right lane after clearing the green-light intersection, \ldots''}};
\fill[white,fill opacity=.85,rounded corners=1pt] (\xMe-.88,\rA-.86) rectangle (\xMe-.03,\rA-.47);
\draw[tzInk,line width=.7pt,line cap=butt] (\xMe-.8,\rA-.78) -- (\xMe-.078,\rA-.78);
\node[inner sep=0] at (\xMe-.44,\rA-.62) {20 m};
\fill[white,fill opacity=.85,rounded corners=1pt] (\xMe-.64,\rB-.86) rectangle (\xMe-.03,\rB-.47);
\draw[tzInk,line width=.7pt,line cap=butt] (\xMe-.56,\rB-.78) -- (\xMe-.095,\rB-.78);
\node[inner sep=0] at (\xMe-.33,\rB-.62) {5 m};
\fill[white,fill opacity=.85,rounded corners=1pt] (\xM+.05,\rA-.87) rectangle (\xM+1.62,\rA-.39);
\draw[tzPath,line width=1.1pt] (\xM+.12,\rA-.51) -- (\xM+.42,\rA-.51);
\node[anchor=west,inner sep=0] at (\xM+.47,\rA-.51) {trajectory};
\draw[tzLane,line width=.9pt,dash pattern=on 2.6pt off 1.5pt] (\xM+.12,\rA-.75) -- (\xM+.42,\rA-.75);
\node[anchor=west,inner sep=0] at (\xM+.47,\rA-.75) {best-fit route};
\node[fill=white,rounded corners=1pt,inner sep=1.5pt,font=\scriptsize\bfseries,anchor=north west,text=tzLane!85!black] at (\xM+.07,\rA+.86) {HD map};
\node[anchor=north west,inner sep=0,font=\scriptsize\bfseries,text=tzLane!85!black] at (\xE+.09,\rA+.86) {Lane-relative evidence};
\node[anchor=west,inner sep=0,text=tzLane!85!black] at (\xE+.09,\rA+.27) {routed lane};
\node[read,anchor=east] at (\xEe-.09,\rA+.53) {ends 0.3 m right};
\node[read,anchor=west] at (\xE+.42,\rA-.53) {starts 2.6 m right};
\node[anchor=west,inner sep=0] at (\xE+.09,\rA-.8) {heading +4\textdegree{} vs.\ route +6\textdegree{}};
\node[read,anchor=east] at (\xEe-.14,\rB+.68) {ends 10.2 m left};
\node[read,anchor=west] at (\xE+.09,\rB-.06) {starts 1.2 m right};
\node[anchor=west,inner sep=0] at (\xE+.09,\rB-.8) {heading +101\textdegree{} vs.\ route 0\textdegree{}};
\node[frame,anchor=north west,minimum width=3.1cm,minimum height=1.86cm,inner sep=0] at (\xF,\rA+\rh) {};
\node[frame,anchor=north west,minimum width=3.1cm,minimum height=1.86cm,inner sep=0,draw=tzBad,fill=tzBad!5] at (\xF,\rB+\rh) {};
\fill[tzOK,rounded corners=1pt] (\xF,\rA+\rh) rectangle (\xF+.07,\rA-\rh);
\fill[tzBad,rounded corners=1pt] (\xF,\rB+\rh) rectangle (\xF+.07,\rB-\rh);
\node[anchor=north west,inner sep=0,font=\footnotesize\bfseries,text=tzOK] at (\xF+.2,\rA+.8) {\textsc{consistent}};
\node[anchor=north west,inner sep=0,font=\footnotesize\bfseries,text=tzBad] at (\xF+.2,\rB+.8) {\textsc{inconsistent}};
\node[anchor=north east,draw=tzOK,text=tzOK,rounded corners=1pt,inner sep=1.5pt,font=\scriptsize\bfseries] at (\xFe-.12,\rA+.83) {no alarm};
\node[anchor=north east,fill=tzBad,text=white,rounded corners=1pt,inner sep=1.5pt,font=\scriptsize\bfseries] at (\xFe-.12,\rB+.83) {alarm};
\node[quote] at (\xF+.2,\rA-.2) {``\ldots offset moves steadily left\ldots settling near the reference lane center\ldots''};
\node[quote] at (\xF+.2,\rB-.2) {``\ldots the route heading change is 0 deg\ldots a full left turn, not a lane change\ldots''};
\end{tikzpicture}
\caption{Using a driving VLA's reasoning as a runtime specification. At each planning step, the VLA emits a chain-of-thought (CoT) and a trajectory. We manually assess CoT reliability (check 1), then monitor whether the trajectory performs the stated maneuver (check 2). A curve can look like a lane change in raw coordinates, so our lane-aware monitor measures the trajectory against the map lane it best follows. It accepts (a), the stated lane change, and flags (b), a left turn into the curb.}
\Description{Two example planning steps flow from camera scene through a driving VLA to a chain-of-thought and a trajectory drawn with its best-fit route on the HD map, then to lane-relative evidence that shows the trajectory as offset from the routed lane, then through a judge to a verdict. In the first, the trajectory moves into the routed lane and settles there; it is consistent and raises no alarm. In the second, the trajectory turns about 100 degrees away from a straight route and ends 10 meters left of it; it is inconsistent and raises an alarm.}
\label{fig:teaser}
\end{teaserfigure}

\maketitle
\pagestyle{plain}


\section{Introduction}
\hypersetup{hidelinks}
Autonomous vehicles (AVs) must operate safely within an unpredictable physical
world, where a single failure can lead to catastrophic consequences, as
illustrated by the Uber automated-test-vehicle pedestrian fatality, the Tesla
Autopilot Mountain View crash, and the Cruise driverless pedestrian-dragging
incident~\cite{ntsb_uber_tempe_2019,ntsb_tesla_mountain_view_2020,nhtsa_cruise_consent_order_2024}.
Navigating this operational complexity is causing a paradigm shift in the
underlying software architecture of AVs. Industry is moving from traditional,
modular pipelines, where isolated machine-learning components handle narrow
tasks such as perception, prediction, planning, or control, toward larger
integrated subsystems and end-to-end multimodal driving models, including
vision-language-action (VLA) models~\cite{chen_e2e_ad_survey_2024,wayve_lingo2_2024,hwang_emma_2025,alpamayo-r1, renz_simlingo_2025,nvidia_alpamayo15_model_card}.
These frontier models ingest raw visual streams, navigation context, and
telemetry sources to produce vehicle trajectories, and have shown promising
performance on challenging and long-tail driving scenarios~\cite{wayve_lingo2_2024,hwang_emma_2025,alpamayo-r1,nvidia_alpamayo15_model_card}.

Yet, this leap in capability magnifies a severe validation and verification
(V\&V) challenge because the multimodal billion-parameter space of modern models
substantially expands the space of difficult-to-interpret failure modes.
Concurrently, the black-box nature of deep neural networks makes rigorous V\&V difficult because model internals are not expressed in the semantic vocabulary of formal driving specifications, such as vehicles, pedestrians, lanes, right-of-way, and temporal traffic interactions~\cite{toledo_sgsm_2024,woodlief_sceneflow_2025}. When an end-to-end model misinterprets a complex intersection, miscalculates the intent of a merging vehicle, and generates an erroneous trajectory, V\&V techniques struggle to assert that the behavior is incorrect given the high-dimensional input and the opacity of the model's internal representations.

The recent integration of reasoning VLAs into autonomous driving introduces a unique opportunity to address this challenge. Beyond merely generating raw trajectory coordinates, these models emit an explicit chain-of-thought (CoT)~\cite{drivecot_2024,tian_drivevlm_2024,hwang_emma_2025,renz_simlingo_2025,alpamayo-r1,nvidia_alpamayo15_model_card}.
These CoTs are not generic language streams; in driving VLAs, they are
intended to be grounded in the physical task of driving, capturing domain
concepts such as lane geometry, traffic rules, agent interactions, and dynamic
constraints. For example, the CoT in Fig.~\ref{fig:teaser}(a), generated by NVIDIA's Alpamayo~1.5 VLA~\cite{nvidia_alpamayo15_model_card}, states a maneuver (change lane left) together with its scene-grounded justification (slower traffic in front and a clear gap).
From a V\&V perspective, such a CoT provides a candidate specification for cross-checking model outputs and flagging impending failures.

This use of CoTs for V\&V rests on two assumptions that we systematically investigate in this paper:
\begin{itemize}
    \item The generated CoT must be \emph{reliable}: grounded in the scene, safe, and logically coherent (Fig.~\ref{fig:teaser}, check 1).
    \item The CoT and the generated trajectory must be comparable for \emph{consistency}: whether the trajectory performs the maneuver the CoT states (Fig.~\ref{fig:teaser}, check 2).
\end{itemize} 

Our first contribution targets the first assumption. In Section~\ref{sec:case-study}, we conduct an assessment of a recent open driving VLA trained to produce CoTs together with trajectories. We curate DriveAlignBench, a specialized dataset from NVIDIA's Alpamayo 1.5 VLA for AVs~\cite{nvidia_alpamayo15_model_card}, containing 150 CoT--trajectory
pairs that we annotate for reliability, CoT--trajectory consistency, and trajectory safety. Our analysis reveals that reliability varies across driving-decision categories.
  Among categories with at least 10 examples, reliability ranges from 30.0\% to
  81.3\%, with 33.3\% of CoTs being unreliable overall. Among the 100 reliable CoTs, the generated trajectory is consistent with the CoT in 74\% of cases.

Our second contribution targets the second assumption by formulating CoT--trajectory consistency as a runtime monitoring problem. In Section~\ref{sec:detectors}, we ask whether the generated trajectory realizes the maneuver asserted by the CoT. This comparison is difficult because CoTs state open-vocabulary, scene-relative maneuvers, such as lane changes, yielding, and gap keeping, whereas trajectories expose only low-level ego motion whose interpretation depends on road geometry and scene context. We therefore design a family of automated monitors that compare CoTs and trajectories under increasingly rich evidence representations. Our evaluation in Section~\ref{sec:experiments} shows that the strongest monitor is one that adds lane-relative evidence to the input of its judge, the model that scores consistency. This \emph{lane-aware} monitor raises $F_1$ for all four judges we evaluate, from $0.56$--$0.62$ with the \emph{ego-frame} monitor, which gives the judge only raw waypoints, to $0.65$--$0.82$, suggesting that evidence representation can improve detection across judges. We also find that, for reliable CoTs, inconsistency is associated with unsafe behavior (27\% of human-labeled inconsistent trajectories are unsafe, against 3\% of consistent ones). Human-labeled inconsistencies account for 7 of the 9 unsafe trajectories in the reliable subset of 100 pairs and 10 of the 15 in a hand-picked challenge set from hazardous scenes, and the lane-aware monitor flags most of those. However, some unsafe trajectories are consistent and false alarms are common, so inconsistency alone is not a complete safety monitor.

\noindent\textbf{Reproducibility:}  
We release DriveAlignBench, our monitoring architecture, and experiments at \url{https://github.com/less-lab-uva/drive-the-thoughts}.

\begin{figure*}[t]
\centering
\definecolor{exInk}{HTML}{23272E}
\definecolor{exOK}{HTML}{1B7837}
\definecolor{exBad}{HTML}{B2182B}
\def\pw{4.27}\def\ph{1.962}\def\sky{.346}
\newcommand{\exbox}[5]{%
  \pgfmathsetmacro\bxa{#1/1110*\pw}\pgfmathsetmacro\bxb{#3/1110*\pw}
  \pgfmathsetmacro\bya{-(#2-90)/510*\ph}\pgfmathsetmacro\byb{-(#4-90)/510*\ph}
  \draw[white,line width=1pt,line join=round] (\bxa,\bya) rectangle (\bxb,\byb);
  \draw[#5,line width=.55pt,line join=round] (\bxa,\bya) rectangle (\bxb,\byb);}
\newcommand{\exinsetw}[3]{%
  \pgfmathsetmacro\ixa{\pw-.09-#1}\pgfmathsetmacro\ixb{\pw-.09}\pgfmathsetmacro\iyb{-.09-1}
  \draw[white,opacity=.75,line width=.8pt] (\bxa,\bya) -- (\ixa,\iyb) (\bxb,\bya) -- (\ixb,\iyb);
  \draw[#3,line width=.3pt] (\bxa,\bya) -- (\ixa,\iyb) (\bxb,\bya) -- (\ixb,\iyb);
  \node[anchor=north east,inner sep=0] at (\ixb,-.09) {\includegraphics[width=#1cm,height=1cm]{figures/#2_inset.jpg}};
  \draw[white,line width=1pt] (\ixa,-.09) rectangle (\ixb,\iyb);
  \draw[#3,line width=.55pt] (\ixa,-.09) rectangle (\ixb,\iyb);}
\newcommand{\exinset}[2]{\exinsetw{1.5}{#1}{#2}}
\newcommand{\exframe}[2]{\node[fill=white,rounded corners=.5pt,inner sep=.8pt,font=\scriptsize,anchor=south west] at (\ixa+#1,\iyb+.05) {#2};}
\newcommand{\expanel}[5]{%
  \begin{scope}[shift={(#1,0)}]
    \begin{scope}
      \clip[rounded corners=2pt] (0,0) rectangle (\pw,-\ph);
      \node[anchor=north west,inner sep=0] at (0,\sky) {\includegraphics[width=\pw cm]{figures/#2.jpg}};
    \end{scope}
    #5
    \draw[#3,line width=.8pt,rounded corners=2pt] (0,0) rectangle (\pw,-\ph);
    \node[fill=white,rounded corners=1pt,inner sep=1.5pt,font=\scriptsize\bfseries,anchor=north west] at (.07,-.07) {#4};
  \end{scope}}
\newcommand{\extext}[5]{%
  \node[anchor=north west,text width=\pw cm,align=left,inner sep=0] at (#1,-\ph-.09) {\itshape ``#3''};
  \node[anchor=north west,text width=\pw cm,align=left,inner sep=0] at (#1,-\ph-.66) {\textcolor{#2}{\bfseries #4:} #5};}
\begin{tikzpicture}[x=1cm,y=1cm,font=\scriptsize,text=exInk,line cap=round,
  every node/.style={execute at begin node={\hyphenpenalty=10000\exhyphenpenalty=10000}}]
\node[anchor=base west,inner sep=0,font=\footnotesize\bfseries,text=exOK] at (0,.2) {Reliable};
\draw[exOK,line width=.6pt] (0,.11) -- (8.66,.11);
\node[anchor=base west,inner sep=0,font=\footnotesize\bfseries,text=exBad] at (8.94,.2) {Unreliable};
\draw[exBad,line width=.6pt] (8.94,.11) -- (17.6,.11);
\expanel{0}{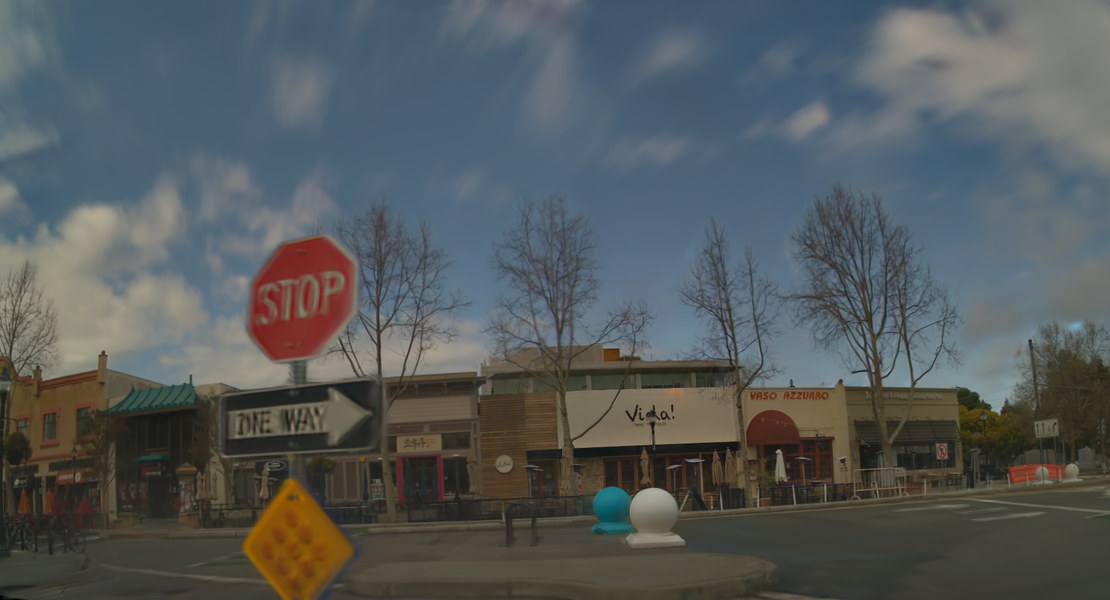}{exOK}{(a) One-way turn}{\exbox{209}{357}{391}{478}{exOK}\exinset{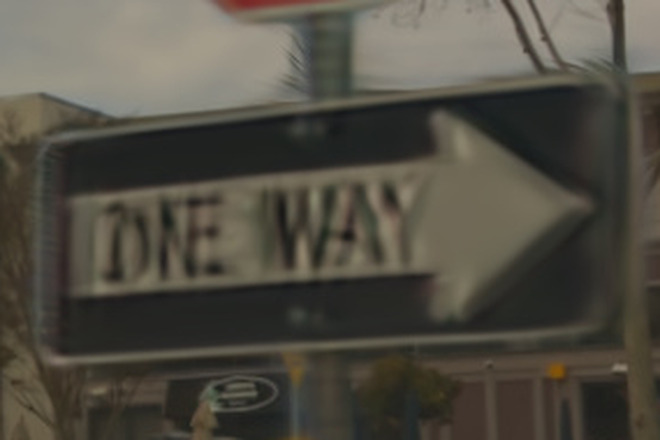}{exOK}}
\expanel{4.39}{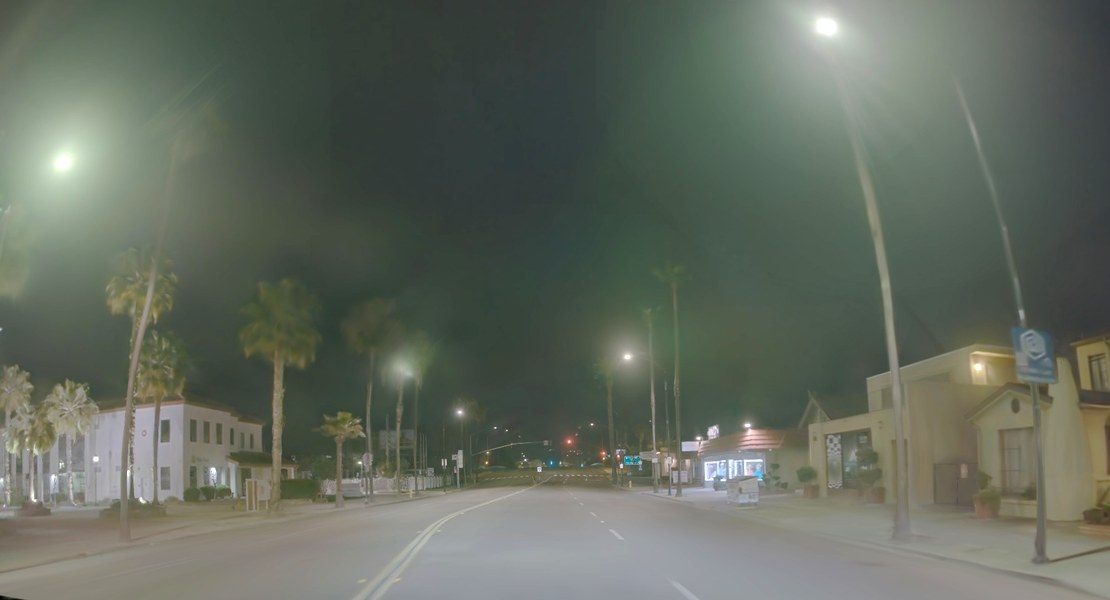}{exOK}{(b) Night stop}{\exbox{552}{426}{630}{478}{exOK}\exinset{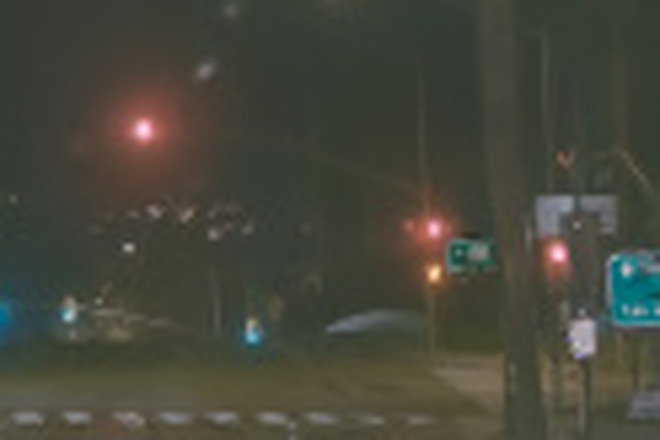}{exOK}}
\expanel{8.94}{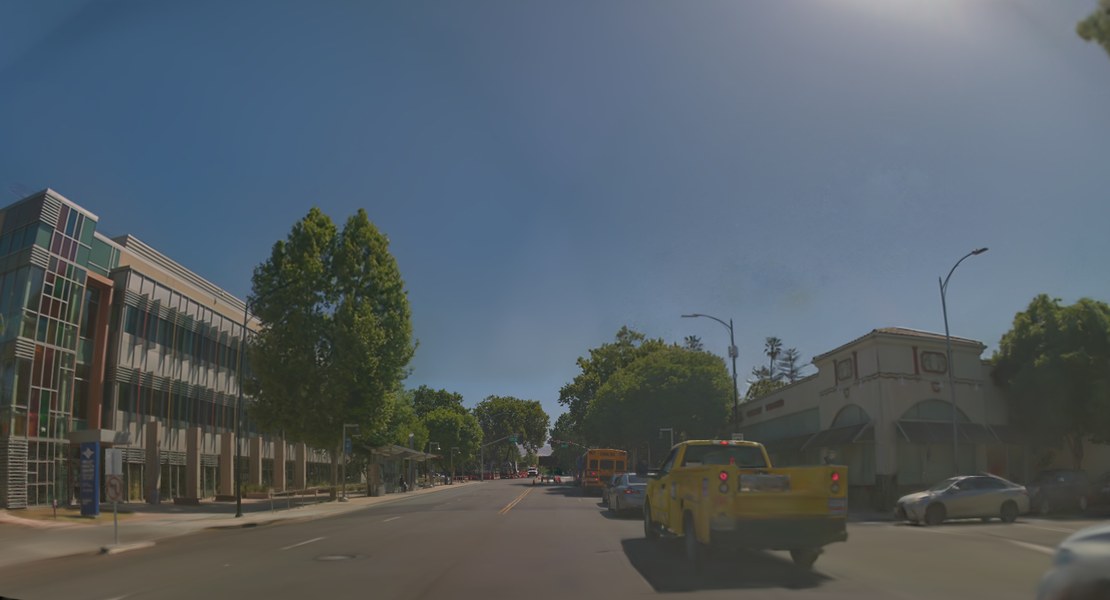}{exBad}{(c) Grounding failure}{\exbox{640}{436}{852}{572}{exBad}}
\expanel{13.33}{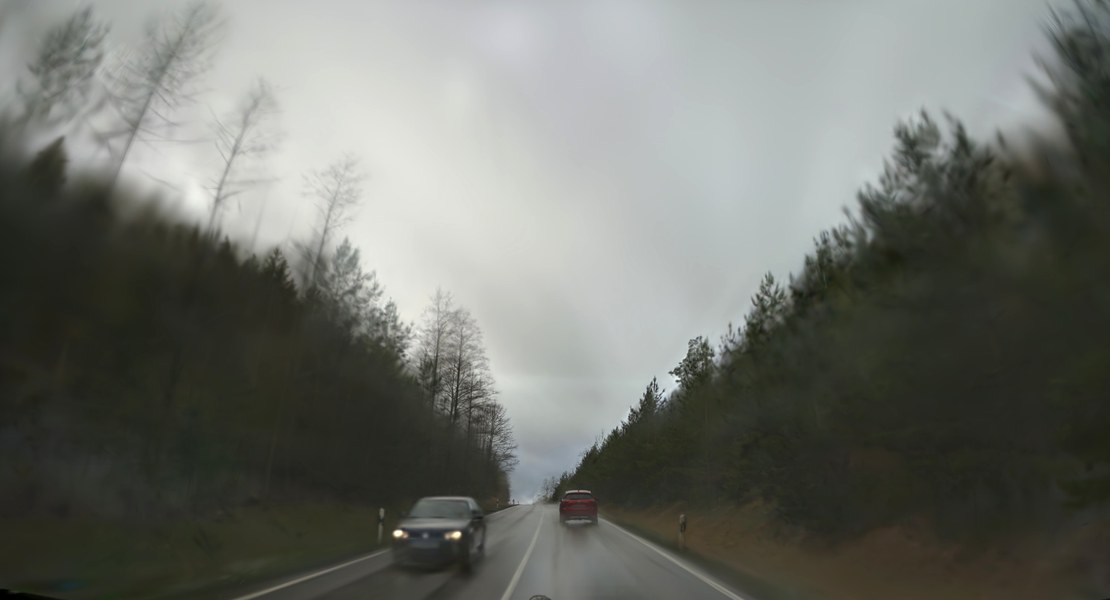}{exBad}{(d) Action-safety failure}{\exbox{390}{494}{490}{572}{exBad}}
\extext{0}{exOK}{Turn right because the one-way sign directs traffic to the right.}{Grounded}{the sign ahead reads ONE WAY with an arrow to the right.}
\extext{4.39}{exOK}{Slow down for the red traffic light ahead since it is red.}{Grounded}{the signal ahead is red.}
\extext{8.94}{exBad}{Keep distance to the lead vehicle since it is directly ahead in our lane.}{Hallucinated lead}{the cited vehicle is in the adjacent lane, not ahead.}
\extext{13.33}{exBad}{Nudge left to overtake the lead vehicle.}{Unsafe maneuver}{nudging left risks hitting the oncoming car in the opposing lane.}
\end{tikzpicture}
\caption{Representative CoTs from DriveAlignBench with the front-camera view
supplied to Alpamayo~1.5.}
\label{fig:cot-examples}
\end{figure*}

\section{Background and Related Work}
Our work focuses on the analysis and monitoring of driving VLA models. More broadly, end-to-end driving systems commonly map sensor and context inputs to a planned future ego trajectory~\cite{chen_e2e_ad_survey_2024}. In reasoning driving VLAs such as LINGO-2 and Alpamayo, the model can condition on visual observations together with route, language, or ego-state context, and can output both text reasoning or commentary and a future trajectory~\cite{wayve_lingo2_2024,nvidia_alpamayo15_model_card}. Let $o_{\le t}$ be the historical observations (including visual observations and historical ego-state) and $\ell$ be the natural-language driving prompt, which is fixed in our experiments. At each \emph{planning step} $t$, one query of the VLA, the model studied here outputs
$(c_t,\tau_t)=f(o_{\le t}, \ell)$
where $c_t$ describes the intended maneuver in text, and $\tau_t = (\mathbf{s}^{*}_{t+1}, \ldots, \mathbf{s}^{*}_{t+H})$ is the generated trajectory whose states carry waypoint positions and headings. Our proposal is to build monitors that check whether $\tau_t$ is consistent with $c_t$.

\subsection{Reasoning in Driving VLAs}
Building on chain-of-thought prompting in language models~\cite{wei_cot_2022}, recent driving VLA systems, such as AdaThinkDrive~\cite{luo_adathinkdrive_2025} and AutoDrive-R$^2$~\cite{yuan_autodriver2_2026}, couple explicit reasoning with action generation~\cite{tang_actiongrounded_survey_2026} to improve trajectory quality, efficiency, or robustness in complex scenarios. Alpamayo models emit Chain-of-Causation reasoning, which we call a CoT, describing driving decisions and causal factors~\cite{alpamayo-r1, nvidia_alpamayo15_model_card}. Alpamayo-R1 further uses reinforcement learning (RL) post-training to improve reasoning quality and enforce reasoning--action consistency~\cite{alpamayo-r1}, and later systems adopt similar training-time signals. WorkDrive rewards agreement between lateral meta-actions (discrete ego-behavior labels) and the predicted trajectory during Group Relative Policy Optimization (GRPO)~\cite{jiang_workdrive_2026}, and Yang et al.\ use a rule-based validator that checks structured CoT fields against the final meta-action as a verifiable GRPO reward~\cite{yang_dualprocess_2026}. Across these systems, however, the CoT is mostly optimized as a means to improve trajectory quality, which leaves two assumptions unverified at runtime: that the reasoning is reliable, and that the trajectory realizes it. Alpamayo-R1, WorkDrive, and Yang et al.\ target the second only during training, through reasoning--action consistency rewards.

\subsection{CoT Faithfulness and Reasoning--Action Consistency}

Prior work on chain-of-thought faithfulness cautions that stated reasoning can be a post-hoc rationalization rather than the model's true decision process~\cite{turpin_2023, lanham_2023}. Recent studies of CoT monitoring in language models add that monitors can detect implicitly induced behavior shifts far less reliably than explicitly instructed ones~\cite{duzan_implicit_2026}, and that poisoned models can act on a backdoor while emitting benign-looking traces, which leads Severi et al.\ to suggest framing CoT monitoring as a consistency check between the trace and the final response~\cite{severi_cotpoisoning_2026}. In robotics, an observable form of this problem is inconsistency between the stated reasoning and the generated action. Foutter et al.\ separate reasoning that is functional, improving task performance, from reasoning that is faithful to the policy's decision process; in a human evaluation on Alpamayo models, RL post-training often improved trajectories while the accompanying reasoning stayed unchanged or degraded, and they train toward faithfulness with a learned critic used as an RL reward~\cite{foutter_faithfulness_2026}. Closest to us, several efforts examine reasoning--action consistency directly. Mayumu et al.\ study Alpamayo-R1 reasoning faithfulness by measuring entity fidelity, action fidelity, reasoning--action consistency, and perturbation sensitivity using keyword extraction, ground-truth checks, and kinematic predicates~\cite{mayumu_2026}. VLADriveBench evaluates ORION and two Alpamayo models in CARLA, combining observational metrics, such as agreement between the CoT's stated longitudinal intent and the action over accelerate, maintain, and decelerate classes, with CoT interventions, and finds that Alpamayo~1.5's reasoning causally shapes its actions while ORION's is epiphenomenal~\cite{nguyen_vladrivebench_2026}. Priyadershi and Frtunikj perturb Alpamayo-R1's sensor inputs and find that trajectories deviate 5.3$\times$ more when its Chain-of-Causation changes, which they argue motivates reasoning-based runtime monitoring~\cite{priyadershi_lostinfog_2026}. Outside driving, Wu et al.\ improve tabletop manipulation robustness by sampling candidate action sequences at runtime, predicting their outcomes in simulation, and using a pretrained vision-language model to select an execution whose outcome aligns with the model's textual plan~\cite{wu_2025}.
We share their premise that reasoning--action consistency, which we call CoT--trajectory consistency when the action is a trajectory, is checkable, but, whereas the driving studies above measure it as an evaluation metric or optimize it during training, we study the challenges of using it for runtime monitoring in driving, where reasoning is often scene-relative while the generated trajectory is expressed as raw waypoints. Our monitor bridges this representation gap rather than assuming that reasoning statements can be directly matched to trajectory predicates, discrete action classes, or candidate actions.

\subsection{Runtime Monitoring and Verification of Driving Policies}

Runtime monitoring offers a complementary way to assure learning-enabled driving systems when offline verification is insufficient. Prior work includes model-based safety envelopes, such as Responsibility-Sensitive Safety, which defines formal kinematic safety constraints for safe planned motion~\cite{shalev_rss_2017}, and hazard monitors that check an end-to-end planner's output trajectory against the predicted motion of surrounding agents and hand control to a fallback controller~\cite{wang_argus_2025}; monitors for learned components, such as out-of-distribution detection and introspective failure prediction~\cite{hendrycks_ood_2017, kuhn_introspective_2020}, misbehavior prediction from autoencoder reconstruction error~\cite{stocco_selforacle_2020}, and recent failure detectors for VLA manipulation policies, which learn failure signals from trajectory-level labels~\cite{park_hideandseek_2026}, score feature-space outliers and inconsistent consecutive action chunks~\cite{seligmann_vlafail_2026}, or probe internal activations for task progress~\cite{bhardwaj_progressprobes_2026}; and semantic runtime monitors that check authored driving properties over relational structures such as scene graphs and temporal scene-flow specifications~\cite{toledo_sgsm_2024, woodlief_sceneflow_2025}. These approaches provide external safety or reliability checks over the planner, perception stack, policy internals, or scene representation. Our work instead monitors whether the VLA's generated trajectory is consistent
with its own CoT description of the intended maneuver.
\section{CoT Reliability Analysis}
\label{sec:case-study}

\begin{figure*}[t]
\centering
\definecolor{relBar}{HTML}{4F55A5}
\definecolor{relMute}{HTML}{6B7280}
\newcommand{\relrow}[4]{%
  \ifnum#3<10 \def\relfill{relBar!36}\def\reltext{relMute}\else\def\relfill{relBar}\def\reltext{black}\fi
  \fill[black!7] (0,#1-.29) rectangle (100,#1+.29);
  \fill[fill=\relfill] (0,#1-.29) rectangle (#4,#1+.29);
  \node[anchor=east,text=\reltext] at (-2,#1) {#2 ($n{=}#3$)};
  \node[anchor=east,text=\reltext] at (130,#1) {#4\%};
}
\newcommand{\relaxis}[1]{%
  \draw[relMute,line width=.35pt] (0,.55) -- (100,.55);
  \foreach \x in {0,50,100}{%
    \draw[relMute,line width=.35pt] (\x,.55) -- (\x,.45);
    \node[anchor=north,inner sep=1pt,text=relMute] at (\x,.43) {\x};}
  \draw[black!80,line width=.5pt,dash pattern=on 2pt off 1.5pt] (66.7,.45) -- (66.7,#1+.42);
  \node[anchor=north,inner sep=1pt] at (66.7,.43) {66.7};
  \node[anchor=north,inner sep=1pt,text=relMute] at (50,-.12) {Reliable CoTs (\%)};
}
\begin{minipage}[t]{0.325\textwidth}
\centering\footnotesize\textbf{(a) Longitudinal decision}\\[2pt]
\begin{tikzpicture}[x=0.021cm,y=0.40cm,font=\scriptsize]
  \relrow{7}{Set speed}{10}{80.0}
  \relrow{6}{Lead follow}{52}{76.9}
  \relrow{5}{Static stop}{16}{62.5}
  \relrow{4}{Gap search}{13}{46.2}
  \relrow{3}{Yield}{8}{37.5}
  \relrow{2}{Overtake accel.}{10}{30.0}
  \relrow{1}{Speed adapt.}{1}{0.0}
  \relaxis{7}
\end{tikzpicture}
\end{minipage}\hfill
\begin{minipage}[t]{0.325\textwidth}
\centering\footnotesize\textbf{(b) Lateral decision}\\[2pt]
\begin{tikzpicture}[x=0.021cm,y=0.40cm,font=\scriptsize]
  \relrow{5}{Lane keep}{75}{81.3}
  \relrow{4}{In-lane nudge}{16}{75.0}
  \relrow{3}{Turn}{12}{75.0}
  \relrow{2}{Merge/split}{5}{40.0}
  \relrow{1}{Lane change}{42}{38.1}
  \relaxis{5}
\end{tikzpicture}
\end{minipage}\hfill
\begin{minipage}[t]{0.325\textwidth}
\centering\footnotesize\textbf{(c) Causal factor}\\[2pt]
\begin{tikzpicture}[x=0.021cm,y=0.40cm,font=\scriptsize]
  \relrow{7}{Road events}{42}{61.9}
  \relrow{6}{Critical objects}{97}{60.8}
  \relrow{5}{Yield/stop control}{10}{60.0}
  \relrow{4}{ODD constraints}{5}{60.0}
  \relrow{3}{Traffic lights}{26}{57.7}
  \relrow{2}{Lane/lanelines}{82}{53.7}
  \relrow{1}{Routing intent}{25}{44.0}
  \relaxis{7}
\end{tikzpicture}
\end{minipage}
\caption{CoT reliability by decision or causal-factor tag. The dashed line marks
overall reliability (66.7\%), and lighter bars mark tags with $n<10$. Panels are
overlapping views of the same 150 pairs: a pair has at most one tag per decision
channel (40 pairs state no longitudinal decision) but may have multiple causal
factors. ODD: operational design domain.}
\label{fig:reliability-by-category}
\end{figure*}

A CoT--trajectory consistency check is meaningful only if the CoT is itself a
valid description of what the vehicle should do. Otherwise, the monitor output becomes difficult to interpret. A trajectory may be consistent with an ungrounded or unsafe CoT, or may be flagged as inconsistent simply because the CoT is unreliable in the first place. 
We therefore first ask when a driving CoT can serve as a \emph{usable specification} for
the monitors introduced in Section~\ref{sec:detectors}.  We study this question
on 150 CoT--trajectory pairs generated by Alpamayo~1.5 and address two research
questions:
\begin{itemize}
  \item \textbf{RQ1:} What proportion of CoTs are grounded in the scene, prescribe
  a safe maneuver, and connect their scene preconditions to that maneuver coherently?
  \item \textbf{RQ2:} For which driving decisions and causal factors are CoTs
  most and least often reliable?
\end{itemize}

\subsection{When Is a CoT a Usable Specification?}
\label{subsec:reliability-criteria}

Let $c$ be a CoT generated from the observation history $o$, and let $m(c)$ be the maneuvers $c$ proposes. We treat $c$ as reliable, i.e., usable as a specification, only when it satisfies three criteria:
\begin{equation}
  \operatorname{Reliable}(c,o)
  =G(c,o)\ \land\ S(m(c),o)\ \land\ L(c,o).
  \label{eq:cot-reliability}
\end{equation}

We view a CoT as a candidate specification because it identifies scene preconditions and uses them to justify a maneuver whose execution should satisfy safety postconditions. We assess CoT reliability with three annotation criteria:

\noindent\textbf{Scene grounding ($G$).}
Every scene precondition used to justify the maneuver must be accurate and supported by the model's visual input. For example, ``stop because the light is red'' requires the relevant signal to be visibly red. We reject claims without sufficient visual support, such as a slower lead vehicle when relative motion is unclear.

\noindent\textbf{Action safety ($S$).}
Every proposed maneuver must be safe in the ground-truth scene underlying $o$. Executing it must not make a collision unavoidable, pose significant risk to other agents or obstacles, leave the road, or violate a safety-related traffic law. For example, stopping for a pedestrian is safe, whereas overtaking into an opposing lane toward an oncoming vehicle is unsafe.

\noindent\textbf{Logical coherence ($L$).}
The CoT must be internally coherent, and its grounded preconditions must plausibly justify the stated maneuver. For example, a slower lead vehicle and a safe target-lane gap can justify a lane change. In contrast, changing into a merging vehicle's lane does not follow from the stated goal of yielding to it.

A CoT is usable as a specification only if all three criteria hold. Failures are non-exclusive, where one CoT may, for example, hallucinate a lane and prescribe an unsafe maneuver into it.

\subsection{Benchmark and Evaluation Protocol}
\label{subsec:methodology}
\noindent\textbf{Collection.}
Alpamayo~1.5 is an open 10B-parameter reasoning VLA that emits a natural-language
CoT followed by a 6.4\,s ego trajectory~\cite{nvidia_alpamayo15_model_card,alpamayo-r1}.
We study Alpamayo~1.5 because it is a widely used open reasoning
VLA, and its model family is the common testbed of concurrent work on driving-CoT
faithfulness~\cite{mayumu_2026,nguyen_vladrivebench_2026,foutter_faithfulness_2026,priyadershi_lostinfog_2026}.
Its released weights and AlpaSim support make every pair reproducible, and its
CoTs follow an explicit decision taxonomy~\cite{alpamayo-r1} that gives our
monitors a well-defined action vocabulary.
We run it in the released AlpaSim configuration~\cite{nvidia_alpasim} on 3D scenes that
NVIDIA NuRec reconstructs from real drives, recording each emitted CoT and trajectory as one pair.
Scenes carry metadata for layout, road type, weather, and traffic density.  For
coverage tracking and reliability analysis, we tag each CoT with GPT-5.5~\cite{openai_gpt55} using
the Alpamayo-R1 longitudinal-decision, lateral-decision, and causal-factor
taxonomy~\cite{alpamayo-r1}, where a causal factor is the scene reason a CoT gives for its decision.  We first select scenes to cover the 40
provided scene-label categories, then favor each rollout (a closed-loop simulation of a scene) with previously unseen
taxonomy tags, and sample randomly after coverage saturates. The resulting
DriveAlignBench includes 150 pairs covering all longitudinal decision categories and five
lateral decision categories; out-of-lane nudge, pull-over/curb approach, and
lateral-maneuver abort are absent because Alpamayo rarely or never produces CoTs
that GPT-5.5 maps to them.

\noindent\textbf{Annotation.}
Two authors apply these criteria using the CoT and scene context (the ego-camera
frames, the video up to planning time, and the ego pose). They do not use the trajectory to ensure
CoT reliability is judged independently of CoT--trajectory consistency. Both
annotators first label a 20-pair calibration set and refine the worked examples;
one annotator then labels the remaining pairs, while the second independently
labels 10 of them. On these double-annotated pairs, the annotators agree on the final
reliable/unreliable label for 9/10 pairs. For each pair, we retain the three
reliability-criterion labels, a short justification for each violation, and the
decision and causal-factor tags.

\begin{figure*}[t]
\centering
\definecolor{mpInk}{HTML}{23272E}
\definecolor{mpMute}{HTML}{6B7280}
\definecolor{mpBand}{HTML}{F4F5F7}
\definecolor{mpRule}{HTML}{C5CAD3}
\definecolor{mpJudge}{HTML}{4F55A5}
\definecolor{mpJudgeF}{HTML}{E8E9F7}
\definecolor{mpLane}{HTML}{14806F}
\definecolor{mpLaneF}{HTML}{DDF2EC}
\definecolor{mpPath}{HTML}{E07B1A}
\definecolor{mpRoad}{HTML}{5B8DB8}
\definecolor{mpEgo}{HTML}{1B7837}
\definecolor{mpBad}{HTML}{B2182B}
\begin{tikzpicture}[x=1cm,y=1cm,font=\scriptsize,text=mpInk,line cap=round,
  flow/.style={-{Stealth[length=1.5mm,width=1.2mm]},line width=.5pt,draw=mpInk!75},
  stage/.style={rounded corners=1.5pt,line width=.45pt,align=center,inner sep=2pt},
  neutral/.style={stage,draw=mpRule,fill=white},
  judge/.style={stage,draw=mpJudge,fill=mpJudgeF,minimum width=2.3cm,minimum height=.82cm},
  road/.style={stage,draw=mpLane,fill=mpLaneF},
  label chip/.style={rounded corners=1pt,draw=mpRule,fill=white,inner sep=2pt,font=\scriptsize\ttfamily},
  evid/.style={rounded corners=2pt,draw=mpRule,line width=.4pt},
  axis/.style={-{Stealth[length=1mm,width=.9mm]},line width=.35pt,draw=mpMute},
  head/.style={font=\scriptsize,text=mpMute},
  ego plot/.pic={
    \draw[axis] (0,0) -- (1.72,0) node[right=-1pt,text=mpMute] {$x$};
    \draw[axis] (0,0) -- (0,.82) node[left=-1pt,pos=.92,text=mpMute] {$y$};
    \draw[mpPath,line width=.75pt,domain=0:1,samples=24,smooth] plot ({1.55*\x},{.7*\x*\x*\x});
    \foreach \t in {.2,.4,.6,.8,1}{\fill[mpPath] ({1.55*\t},{.7*\t*\t*\t}) circle (1.15pt);}
    \node[anchor=north west,inner sep=0] at (.09,.8) {$\phi_{\mathrm{ego}}(\tau)$};
  },
  lane plot/.pic={
    \draw[mpRoad,line width=.5pt] (0,.09) -- (1.55,.09) (0,.47) -- (1.55,.47);
    \draw[mpRoad,line width=.4pt,dash pattern=on 2pt off 1.5pt] (0,.28) -- (1.55,.28);
    \draw[axis] (0,0) -- (1.72,0) node[right=-1pt,text=mpMute] {$l_k$};
    \draw[axis] (0,0) -- (0,.62) node[left=-1pt,pos=.9,text=mpMute] {$e_k$};
    \draw[mpPath,line width=.75pt,domain=0:1,samples=24,smooth] plot ({1.55*\x},{.28+.03*sin(190*\x)});
    \foreach \t in {.2,.4,.6,.8,1}{\fill[mpPath] ({1.55*\t},{.28+.03*sin(190*\t)}) circle (1.15pt);}
    \node[anchor=south,inner sep=0] at (.82,.63) {$\phi_{\mathrm{lane}}(\tau,\mathrm{route})$};
  }]
\def\rA{-1.13}\def\rB{-2.67}\def\rC{-4.21}
\foreach \y in {\rA,\rB,\rC}{\fill[mpBand,rounded corners=2pt] (2.3,\y+.72) rectangle (15.68,\y-.72);}
\node[head] at (6.95,-.2) {Evidence};
\node[head] at (11.6,-.2) {Comparison};
\node[head] at (14.1,-.2) {Threshold};
\node[anchor=west,align=left,inner sep=0] at (2.42,\rA) {{\footnotesize\bfseries Rule-based}\\[1pt]{\color{mpMute}closed}\\{\color{mpMute}vocabulary}};
\node[anchor=west,align=left,inner sep=0] at (2.42,\rB) {{\footnotesize\bfseries Ego-frame}\\[1pt]{\color{mpMute}open}\\{\color{mpMute}vocabulary (C1)}};
\node[anchor=west,align=left,inner sep=0] at (2.42,\rC) {{\footnotesize\bfseries Lane-aware}\\[1pt]{\color{mpMute}lane-relative}\\{\color{mpMute}evidence (C2)}};
\draw[mpRule,line width=.45pt,fill=white,rounded corners=2pt] (0,0) rectangle (1.98,-4.93);
\node[font=\footnotesize\bfseries] at (.99,-.26) {VLA output};
\node[anchor=north,text width=1.72cm,align=center,inner sep=0] at (.99,-.62) {CoT $c$\\[2pt]{\itshape ``Keep lane since the lane is clear ahead''}};
\node at (.99,-2.32) {Trajectory $\tau$};
\begin{scope}[shift={(.2,-3.72)}]
  \draw[mpRoad,line width=.5pt,domain=0:1,samples=24,smooth] plot ({1.55*\x},{.72*\x*\x*\x+.2});
  \draw[mpRoad,line width=.5pt,domain=0:1,samples=24,smooth] plot ({1.55*\x},{.72*\x*\x*\x-.2});
  \draw[mpPath,line width=.8pt,domain=0:1,samples=24,smooth] plot ({1.55*\x},{.72*\x*\x*\x});
  \fill[mpEgo] (-.1,-.07) rectangle (.14,.07);
\end{scope}
\node[align=center,text=mpMute] at (.99,-4.3) {follows a\\left curve};
\foreach \y in {\rA,\rB,\rC}{\draw[flow] (1.98,\y) -- (2.3,\y);}
\node[evid,minimum width=5.9cm,minimum height=1.28cm,anchor=west] (eA) at (4.1,\rA) {};
\node[neutral,minimum height=.42cm,minimum width=.62cm] (pc) at (4.92,\rA+.29) {$P_c$};
\node[neutral,minimum height=.42cm,minimum width=.62cm] (pt) at (4.92,\rA-.29) {$P_\tau$};
\node[anchor=west,inner sep=1pt] (cA) at (4.2,\rA+.29) {$c$};
\node[anchor=west,inner sep=1pt] (tA) at (4.2,\rA-.29) {$\tau$};
\draw[flow] (cA) -- (pc); \draw[flow] (tA) -- (pt);
\node[label chip,anchor=west] (ac) at (5.62,\rA+.29) {go\_straight};
\node[label chip,anchor=west] (at) at (5.62,\rA-.29) {steer\_left};
\draw[flow] (pc) -- (ac); \draw[flow] (pt) -- (at);
\node[anchor=west,text=mpMute] at (7.42,\rA+.29) {$\mathbf{a}_c$: phrase dictionary};
\node[anchor=west,text=mpMute] at (7.42,\rA-.29) {$\mathbf{a}_\tau$: kinematic thresholds};
\node[neutral,minimum width=2.3cm,minimum height=.82cm] (m) at (11.6,\rA) {{\bfseries Match}\\exact labels, in order};
\draw[flow] (eA.east) -- (m);
\node[evid,minimum width=5.9cm,minimum height=1.28cm,anchor=west] (eB) at (4.1,\rB) {};
\node[anchor=west,inner sep=1pt] (cB) at (4.2,\rB) {$c$\,\textbf{+}};
\pic at (4.72,\rB-.42) {ego plot};
\node[anchor=west,align=left,inner sep=0,text width=3.0cm] at (6.92,\rB) {Raw ego-frame waypoints.\\{\color{mpMute}A curve reads as a lateral shift.}};
\node[judge] (jB) at (11.6,\rB) {{\bfseries Judge}\\prompt + rubric};
\draw[flow] (eB.east) -- (jB);
\node[neutral,minimum height=.46cm,minimum width=1cm] (tB) at (14.1,\rB) {$s\le\theta$};
\draw[flow] (jB) -- node[above=-1pt] {$s,r$} (tB);
\node[evid,minimum width=5.9cm,minimum height=1.28cm,anchor=west] (eC) at (4.1,\rC) {};
\node[anchor=west,inner sep=1pt] at (4.2,\rC) {$c$\,\textbf{+}};
\pic at (4.72,\rC-.42) {ego plot};
\fill[mpLaneF!55,rounded corners=1.5pt] (6.44,\rC-.59) rectangle (9.94,\rC+.59);
\node[road,minimum height=.36cm,minimum width=1.06cm,fill=white] (hd) at (7.03,\rC+.29) {HD map};
\node[road,minimum height=.36cm,minimum width=1.06cm,fill=white] (rt) at (7.03,\rC-.29) {select route};
\draw[flow,draw=mpLane] (hd) -- (rt);
\pic at (7.98,\rC-.5) {lane plot};
\draw[flow,draw=mpLane] (rt.east) -- (7.96,\rC-.29);
\node[judge] (jC) at (11.6,\rC) {{\bfseries Judge}\\prompt + rubric};
\draw[flow] (eC.east) -- (jC);
\node[neutral,minimum height=.46cm,minimum width=1cm] (tC) at (14.1,\rC) {$s\le\theta$};
\draw[flow] (jC) -- node[above=-1pt] {$s,r$} (tC);
\draw[mpJudge,line width=.45pt,dash pattern=on 2pt off 1.5pt] (jB.south) -- node[right=1pt,text=mpJudge] {same judge} (jC.north);
\draw[mpRule,line width=.45pt,fill=white,rounded corners=2pt] (16.05,0) rectangle (17.6,-4.93);
\node[font=\footnotesize\bfseries] at (16.83,-.26) {Verdict};
\node[font=\Large] at (16.83,-2.2) {$\hat{y}$};
\node[align=center] at (16.83,-2.95) {\textcolor{mpEgo}{\textsc{consistent}}\\[1pt]or\\[1pt]\textcolor{mpBad}{\textsc{inconsistent}}};
\draw[flow] (m) -- (16.05,\rA);
\draw[flow] (tB) -- (16.05,\rB);
\draw[flow] (tC) -- (16.05,\rC);
\end{tikzpicture}
\caption{Progression of the three consistency monitors on a lane-keeping example:
label matching over a shared closed vocabulary, a judge over the CoT and raw
ego-frame waypoints, and the lane-aware monitor, which keeps the same judge but adds lane-relative
evidence, so that ordinary curve following no longer reads as a lateral maneuver.}
\label{fig:monitor_progression}
\end{figure*}

\subsection{Reliability Findings}
\label{subsec:results}

\noindent\textbf{Overall reliability (RQ1).}
100/150 CoTs (66.7\%) satisfy all three criteria. Of the 50 unreliable
CoTs, 43 (86.0\%) contain a grounding failure, 7 (14.0\%) prescribe an unsafe
maneuver, and 6 (12.0\%) are logically incoherent; 6 fail two criteria.  Grounding is therefore
the dominant obstacle to treating a CoT as a specification. Fig.~\ref{fig:cot-examples} shows examples on both reliable and unreliable CoTs. Because sampling
targets scene and behavior coverage, the rate only characterizes this
benchmark.

\noindent\textbf{Which CoTs are reliable (RQ2).}
Fig.~\ref{fig:reliability-by-category} shows the benchmark's reliability by driving decision and causal factor. Set-speed tracking
(8/10), lead following (40/52), and lane keeping (61/75) are the most reliable
well-represented decision categories. Reliability falls sharply for lane
changes (16/42), gap-searching (6/13), yielding (3/8), and acceleration for
overtaking (3/10).
Among causal factors, routing intent (11/25) and lane/laneline reasoning (44/82)
are least reliable.

Overall, CoTs are more reliable for routine, directly observable driving behavior, and less reliable when they must infer lane topology, routing intent, or whether a spatial or temporal gap is available. For these weaker categories, Alpamayo's CoTs require additional reliability
screening before being used as specifications.

\section{Monitoring for Inconsistency Detection}
\label{sec:detectors}

For reliable CoTs (Section~\ref{sec:case-study}), the remaining runtime
monitoring task is to check whether the generated trajectory realizes the
stated maneuver. Given a reliable CoT and a generated trajectory, a monitor should
flag cases where the plan contradicts, omits, or materially deviates from the
maneuver stated in the CoT.

\subsection{Problem Formulation}

\noindent\textbf{Challenges.}
Monitoring CoT--trajectory consistency is difficult because the two artifacts
describe driving behavior at different levels of abstraction. The CoT states
a maneuver in natural language, while the trajectory records future ego motion as
geometry and kinematics. A monitor must bridge this gap along two dimensions.

\emph{(C1) Open-vocabulary ego-maneuver extraction.}
The CoT $c$ is a free-form natural-language artifact rather than a closed-set action label. The same ego maneuver may be expressed in many ways, such as \emph{move over}, \emph{merge right}, or \emph{position for the upcoming turn}. A monitor must therefore recover the intended ego maneuver from open-vocabulary CoTs.

\emph{(C2) Semantic gap between CoT and trajectory.}
The maneuvers stated in $c$ are often scene-relative; they reference lanes, map
elements, and other agents. The trajectory $\tau$, by contrast, is typically a
sequence of waypoints that yield speed, acceleration, and heading rate but not the
entities $c$ refers to. Whether a stated maneuver is realized therefore often
cannot be decided from ego motion alone: \emph{keep lane} requires the lane
geometry, \emph{stop before the crosswalk} the crosswalk's position, and
\emph{move left to avoid an obstacle} the obstacle's location.

We formalize this comparison as follows. A consistency monitor $M$ takes a CoT
$c$ and a representation of the generated trajectory produced by a transform
$\phi$, and returns a \emph{verdict}
\[
  \hat{y} = M\bigl(c,\;\phi(\tau,\xi)\bigr),\,
  \hat{y}\in\{\textsc{consistent},\textsc{inconsistent}\}.
\]
Here, $\tau$ denotes the generated trajectory, and $\xi$ denotes optional
context; we write $\phi(\tau)$ when $\xi=\varnothing$. We call the
output of $\phi$ the \emph{evidence} the monitor reads, and its form the
\emph{evidence representation}.

A pair is inconsistent if the generated trajectory contradicts, omits, or
materially deviates from the stated maneuver, meaning it no longer performs that
maneuver. For example, the trajectory keeps the lane when the CoT specifies a lane change, merely
decelerates when the CoT calls for a complete stop, or accelerates through a
conflict zone when the CoT states a yield. The monitor raises an
\emph{alarm} when it returns \textsc{inconsistent}; an alarm on an
inconsistent pair is a correct detection, and an alarm on a consistent
pair is a false alarm.

\subsection{Monitoring Approaches}

Our central design question is which evidence representation makes $c$ and
$\tau$ comparable. We therefore design three monitors that progressively enrich
the evidence available for consistency checking
(Fig.~\ref{fig:monitor_progression}). The rule-based monitor gives an
interpretable starting point based on discrete label matching. The ego-frame
monitor replaces fixed phrase matching with a judge, usually
a large language model, over raw ego-frame waypoints, addressing open-vocabulary maneuver extraction (C1).
The lane-aware monitor further adds lane-relative evidence, addressing
the semantic gap (C2) for maneuvers such as lane keeping, nudging, and lane
changes. We call these last two \emph{judge-based monitors}.

\subsubsection{Rule-Based Label-Matching Monitor}
The rule-based monitor maps both artifacts into a shared closed-set vocabulary
$A$ of ego behaviors and then performs exact label matching. A CoT parser $P_c$
and a trajectory parser $P_\tau$ produce action sequences
$\mathbf{a}_c=P_c(c)\in A^{*}$ and $\mathbf{a}_\tau=P_\tau(\tau)\in A^{*}$, and
the monitor returns $\operatorname{Match}(\mathbf{a}_c,\mathbf{a}_\tau)$:
\textsc{consistent} only when every maneuver stated in the CoT appears, in order,
as the same trajectory action label, and \textsc{inconsistent} otherwise.

\noindent\textit{Alpamayo instantiation.}
We use Alpamayo-R1's longitudinal and lateral meta-action vocabulary as $A$,
matching its RL post-training alignment between generated reasoning and
trajectory-derived ego behavior. We instantiate $P_c$ as a dictionary-based
phrase-to-action parser, with a total of 200 phrases mined from a development
split disjoint from the evaluation set and augmented with paraphrases, and
$P_\tau$ as a kinematic parser that thresholds per-segment speed, acceleration,
and curvature into the same action channels.

\subsubsection{Ego-Frame Monitor}
To address open-vocabulary CoT maneuvers (C1), the ego-frame monitor replaces closed-set
phrase matching with a prompted judge over the CoT and ego-frame trajectory
evidence. We query a judge with a prompt template instantiated with the
CoT $c$ and the ego-frame trajectory $\phi_{\text{ego}}(\tau)$. The template
contains a coordinate-frame legend, few-shot examples, and a scoring rubric.
The judge returns a consistency score $s \in \{1,\dots,5\}$, from 1 (the
trajectory contradicts the CoT) to 5 (clear consistency), and, optionally, with a
short rationale $r$ that we use for error analysis:
\begin{equation}
  (s, r) = \mathrm{Judge}\big(\mathrm{prompt}(c, \phi_{\text{ego}}(\tau))\big).
\end{equation}
$M_{\mathrm{ego}}(c, \phi_{\mathrm{ego}}(\tau))$ returns \textsc{inconsistent} if
$s\le\theta$ and \textsc{consistent} otherwise.

\noindent\textit{Alpamayo instantiation.}
We set $\theta=2$, the rubric's boundary between mismatch (scores 1--2) and alignment (3--5). To form $\phi_{\mathrm{ego}}(\tau)$, we derive ego-motion
quantities from the original $10$\,Hz trajectory and downsample to approximately
$1.25$\,Hz by retaining one waypoint every $0.8$\,s, which keeps the maneuver-level
shape while reducing prompt length and the number of values the judge must process. Each retained waypoint contains ego position
$(x,y)$, speed $v$, lateral velocity $\dot{y}$, and longitudinal acceleration $a$;
the prompt also includes duration, path length, final longitudinal and lateral
displacement, lateral-deviation range, and mean and maximum speed.

\subsubsection{Lane-Aware Monitor} 
\label{sec:fllm_example}
The lane-aware monitor $M_{\mathrm{lane}}$ augments
$\phi_{\mathrm{ego}}(\tau)$ with lane-relative evidence. This targets CoT
maneuvers defined relative to the lane: ``keep lane,'' ``nudge left,'' and ``change
lanes to the right'' are better measured by offset from the lane center than by
ego-frame lateral displacement. Ego-frame waypoints measure lateral motion against
a straight ray along the ego's initial heading and carry no lane geometry, so
the judge must infer how much of that motion comes from the road and how much comes from
the maneuver. Without lane geometry, ordinary curve following can look
like a lateral maneuver and cause a false positive.

Fig.~\ref{fig:fllm-recovery} shows an example of such a false positive. The CoT says
``Nudge left due to the trash bin encroaching into our lane,'' but the trajectory shows rightward lateral displacement in the ego frame. With only this evidence, the
judge may read the trajectory as moving the opposite way and flag the pair as inconsistent.
However, the routed lane actually bends right, and the trajectory moves from $0.96$\,m right of the lane center to $0.54$\,m left of it. With this evidence, it is clearer that the ego nudged left while following the road, making the judge more likely to correctly accept the pair.

Starting from a high-definition (HD) map\footnote{We assume that access to HD maps, such as those used extensively by AVs as part of their driving stack, is available~\cite{li2022hdmapnet}.}, we retrieve connected route candidates from the ego pose and select
\begin{equation*}
\mathrm{route} = \arg\min_{\mathrm{route}_j} \sum_{k} e_{k,j}^{2},
\end{equation*}
where $e_{k,j}$ is the signed lateral offset of waypoint $k\in\{0,\dots,H\}$ from candidate
$\mathrm{route}_j$, and $k=0$ is the ego's current position. We call the minimizer the \emph{best-fit route}, and the lane it follows the
\emph{routed lane}. We compute
$\phi_{\mathrm{lane}}(\tau,\mathrm{route})$ by projecting all waypoints onto this
route. The resulting offsets expose whether the ego stays near the lane
center, drifts, nudges, or changes lanes.

\noindent\textit{Alpamayo instantiation.} $M_{\mathrm{lane}}$ queries
AlpaSim for the local HD map, builds a directed lane graph from lane-center
polylines, and enumerates route candidates from the ego's initial pose, scoring
them over the full $10$\,Hz trajectory ($H=64$). Its
prompt contains the raw ego-frame evidence plus longitudinal progress $l_k$, signed
lateral offset $e_k$, offset change $\Delta e_k=e_k-e_0$, and lateral velocity
$\dot e_k$, along with the heading change of the
trajectory and of the route. We call these added quantities \emph{lane-relative
evidence}. The vocabulary, phrase dictionary,
thresholds, prompts, and implementation details of all three monitors are in the
public repository.

 \section{Evaluating the Monitoring Approaches}
\label{sec:experiments}

We organize the evaluation around three research questions.
\begin{itemize}
    \item \textbf{RQ3:} How accurately do the monitoring approaches  distinguish
    \textsc{consistent} from \textsc{inconsistent} CoT--trajectory pairs on
    reliable CoTs, and how does performance vary with the judge and the
    evidence representation?
    \item \textbf{RQ4:} Is CoT--trajectory inconsistency associated with trajectory safety, and how well does it function as a safety signal compared with safety-monitor baselines?
    \item \textbf{RQ5:} Can the monitors operate at runtime? Specifically, how
    does each monitor's inference latency compare with the $10$\,Hz per-plan rate?
\end{itemize}

\subsection{Experimental Setup}
\label{subsec:setup}
\paragraph{Objects of Evaluation}
Following the protocol in Section~\ref{sec:case-study}, we annotate each pair with two additional binary labels: a CoT--trajectory consistency label and a trajectory-safety label. The \emph{consistency} label records whether $\tau$ realizes the maneuver asserted in $c$, in the sense defined in Section~\ref{sec:detectors}. To ensure accuracy is not confounded by
ungrounded, unsafe-to-follow, or incoherent CoTs whose consistency cannot be
meaningfully checked, we evaluate the consistency monitors only on the
reliable subset of the benchmark in Section~\ref{sec:case-study}: the
$N=100$ annotated pairs whose CoTs are reliable. Each annotator is given the CoT text $c$, the trajectory $\tau$ (including its bird's-eye-view rendering, raw waypoints, and kinematic summary), the relevant ego-camera frame, and the full video and metadata for context. The \emph{trajectory-safety} label, assigned independently of $c$, marks whether $\tau$, if executed,
would create a \emph{hazard}---an outcome that the action-safety criterion (Section~\ref{subsec:reliability-criteria}) rules out, applied here to the trajectory rather than to the CoT.
Fig.~\ref{fig:teaser} shows a consistent, safe case (a) and an inconsistent, unsafe one (b).
Of the $100$ pairs in the reliable subset, $26$ are \textsc{inconsistent} and $74$ \textsc{consistent}, and $9$ are \textsc{unsafe} and $91$ \textsc{safe}. Including the unreliable subset, we have a total of $17$ \textsc{unsafe} pairs.
Agreement among annotators on the double-annotated pairs of Section~\ref{subsec:methodology} is 10/10 for both consistency and safety labels.

\begin{figure*}[t]
\centering
\definecolor{fgInk}{HTML}{23272E}
\definecolor{fgMute}{HTML}{6B7280}
\definecolor{fgRule}{HTML}{C5CAD3}
\definecolor{fgOK}{HTML}{1B7837}
\definecolor{fgBad}{HTML}{B2182B}
\definecolor{fgRoute}{HTML}{14806F}
\definecolor{fgPath}{HTML}{E07B1A}
\def\ph{3.6}\def\pw{5.2}
\newcommand{\bevpanel}[9]{%
  \begin{scope}[shift={(#1,0)}]
    \node[anchor=north west,inner sep=0] at (0,0) {\includegraphics[width=\pw cm,height=\ph cm]{figures/#2}};
    \draw[fgRule,line width=.45pt,rounded corners=2pt] (0,0) rectangle (\pw,-\ph);
    \node[chip] at (.07,-.07) {#3};
    \fill[white,fill opacity=.88,rounded corners=1pt] (.06,-\ph+.06) rectangle (1.6,-\ph+.58);
    \draw[fgPath,line width=1.1pt] (.13,-\ph+.44) -- (.41,-\ph+.44);
    \node[anchor=west,inner sep=0] at (.46,-\ph+.44) {trajectory $\tau$};
    \draw[#4,line width=.9pt,dash pattern=on 2.6pt off 1.5pt] (.13,-\ph+.2) -- (.41,-\ph+.2);
    \node[anchor=west,inner sep=0] at (.46,-\ph+.2) {#5};
    \draw[fgInk,line width=.7pt,line cap=butt] (\pw-.62,-\ph+.13) -- (\pw-.1335,-\ph+.13);
    \node[inner sep=0,anchor=south] at (\pw-.377,-\ph+.15) {10 m};
    \node[anchor=north west,inner sep=1.5pt,align=left,text=#4] at (.07,-.5) {start #7\\\textbf{final #6}};
    \node[anchor=west,inner sep=0] at (0,-\ph-.2) {#8};
    \node[fill=#9,rounded corners=1pt,inner sep=1.5pt,font=\scriptsize\bfseries,text=white,anchor=east] at (\pw,-\ph-.2) {\bevchip};
  \end{scope}}
\begin{tikzpicture}[x=1cm,y=1cm,font=\scriptsize,text=fgInk,line cap=round,
  chip/.style={fill=white,rounded corners=1pt,inner sep=1.5pt,font=\scriptsize\bfseries,anchor=north west}]
\begin{scope}
  \clip[rounded corners=2pt] (0,0) rectangle (6.8,-\ph);
  \node[anchor=north west,inner sep=0] at (0,0) {\includegraphics[width=6.8cm,height=\ph cm]{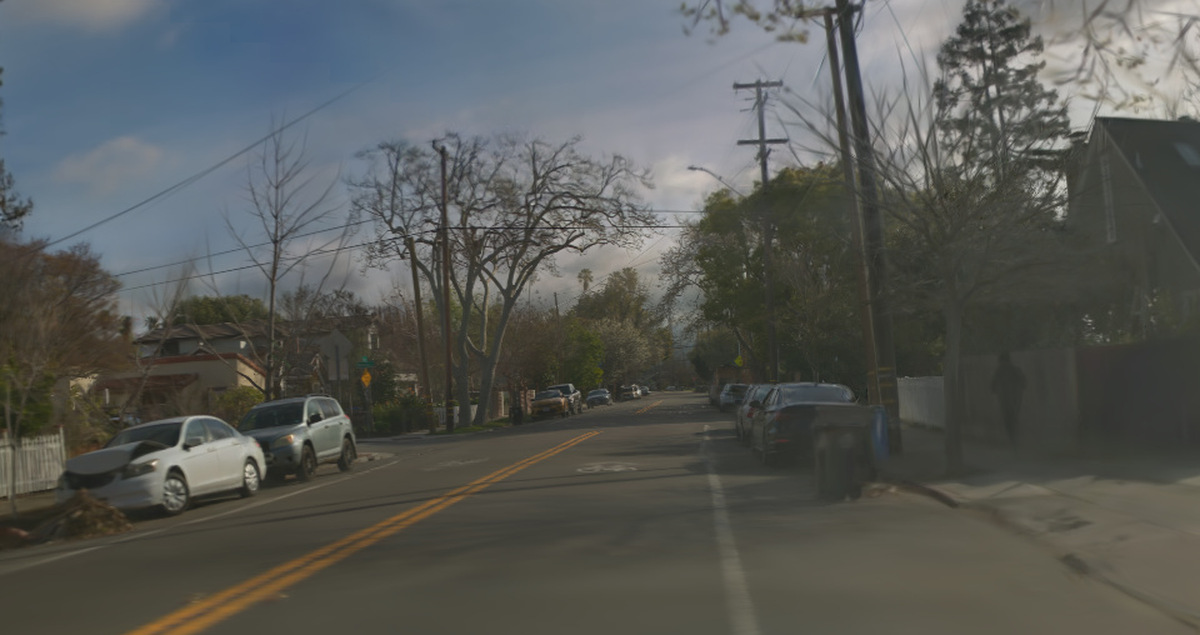}};
\end{scope}
\draw[white,line width=1pt] (4.548,-2.232) rectangle (5.1,-2.87);
\draw[fgInk,line width=.55pt] (4.548,-2.232) rectangle (5.1,-2.87);
\draw[fgRule,line width=.45pt,rounded corners=2pt] (0,0) rectangle (6.8,-\ph);
\node[chip] at (.07,-.07) {(a) Scene};
\node[anchor=west,inner sep=0,font=\scriptsize\itshape] at (0,-\ph-.2) {\textbf{\emph{CoT}}: ``Nudge left due to the trash bin encroaching into our lane.''};
\def\bevchip{wrong}
\bevpanel{7.0}{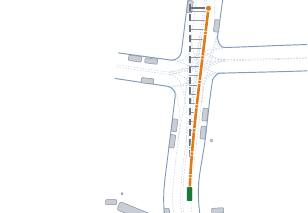}{(b) Raw ego frame $\phi_{\mathrm{ego}}(\tau)$}{fgMute}{ego ray}{$-6.58$ m}{$0.00$ m}{Ego-frame: $1/5$ \textcolor{fgBad}{\textsc{inconsistent}}}{fgBad}
\def\bevchip{correct}
\bevpanel{12.4}{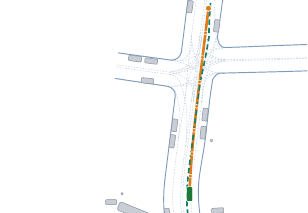}{(c) Lane-relative $\phi_{\mathrm{lane}}(\tau,\mathrm{route})$}{fgRoute}{route}{$+0.54$ m}{$-0.96$ m}{Lane-aware: $4/5$ \textcolor{fgOK}{\textsc{consistent}}}{fgOK}
\end{tikzpicture}
\caption{The same trajectory $\tau$ of a consistent pair, plotted against two
references: the ray along the ego's initial heading (b) and the center of the routed
lane (c). Readouts give the signed lateral offset of $\tau$ from each reference
($+$ = left) at the first and last waypoints the judge reads ($0$ and $6.4$\,s). Against
the ego ray, $\tau$ ends $6.58$\,m to the right, which appears to contradict the stated
left nudge. Against the route, which itself bends right, $\tau$ moves from
$0.96$\,m right of center to $0.54$\,m left, which matches a left nudge. When we ran both
monitors on this pair (judge: Opus 5.5), the ego-frame monitor, which sees only the
evidence in (b), flagged it as inconsistent, while the lane-aware monitor, which also receives (c),
correctly rated it consistent.}
\label{fig:fllm-recovery}
\end{figure*}

\textit{Challenge set.} Because of the limited number of unsafe trajectories, we also evaluate a separate, hand-picked challenge set for RQ4. We ran Alpamayo~1.5 rollouts in AlpaSim on $117$ hazardous or unusual scenes, drawing from NVIDIA's out-of-distribution reasoning~\cite{nvidia_physicalai_av_dataset}, CASCADE~\cite{nvidia_cascade_dataset}, and NuRec~\cite{nvidia_nurec_dataset} datasets. From these, we identified $19$ pairs featuring a reliable CoT and an unsafe trajectory. A second annotator agreed on $13/19$ reliability labels, and on $14/18$ consistency and $17/18$ trajectory-safety labels (excluding unsure cases). Following adjudication by two annotators, $15$ pairs remain, all originating outside the original benchmark scenes. Challenge set contains $10$ \textsc{inconsistent} and $5$ \textsc{consistent} pairs.

\paragraph{Treatments}
We evaluate the three monitors of Section~\ref{sec:detectors}. We run each judge-based monitor with each of the four judges, giving eight
\emph{configurations} (one monitor paired with one judge); the rule-based
monitor, which has no judge, forms a ninth. Every configuration scores all
pairs. The judges are: \textbf{Qwen3.5-4B-FP8}, a small open-weight
model~\cite{redhat_qwen35_4b_fp8}; \textbf{DeepSeek V4.1 Flash 1017}, a large
open-weight model~\cite{deepseek_v41_flash}; \textbf{Claude Opus 5.5}, a
proprietary frontier model~\cite{anthropic_claude_opus55}; and
\textbf{Jev 1.14}, a proprietary decision model that reads a natural-language
prompt and returns a probability for each answer option instead of generating
text~\cite{typesafe_jev}. We call the other three \emph{generative judges}.
Qwen3.5-4B ran locally; the remaining three judges were queried through hosted endpoints. Qwen3.5-4B was queried at temperature $1.0$, top-$p$ $0.95$, top-$k$ $20$, and presence penalty $1.5$. DeepSeek V4.1 Flash and Opus 5.5 were queried at their default reasoning effort setting (\texttt{high}).
We repeated each configuration that uses a judge three times (with $\texttt{seed}\in\{42,43,44\}$ for Qwen3.5-4B and three unseeded runs for the hosted endpoints) to account for non-determinism and mitigate run-to-run variance. Because Jev outputs a probability for each answer option, we adapted the prompt to treat each consistency score ($1$--$5$) as a candidate option. We select the most probable option as the score $s$ (breaking ties toward the lower score); unlike the generative judges, Jev does not produce a rationale $r$.

For RQ4, we also compare against two safety-monitor baselines that flag unsafe
behavior without reading the CoT (settings in Appendix~\ref{app:baselines}):
\begin{itemize}
  \item \textbf{SelfOracle}~\cite{stocco_selforacle_2020}, a black-box misbehavior
  predictor that flags anomalously high autoencoder reconstruction error on camera
  inputs. We adapt the released single-layer (SAE) and variational (VAE)
  autoencoders to AlpaSim camera frames, training each across three seeds on 125,969
  replay images from 630 scenes disjoint from the benchmark, with an alarm threshold at
  $\epsilon=0.05$.
  \item \textbf{\baselineB}, a self-implemented proximity monitor based on the
  two-second following rule, similar to the reachability-based region of
  PhysCov~\cite{hildebrandt_physcov_2023}. It raises an alarm if any other road
  user's bounding box overlaps the trapezoidal region the ego could reach within the
  next $2$\,s.
\end{itemize}

\begin{table}[t]
\caption{Inconsistency detection on the reliable subset ($100$: $26$
\textsc{inconsistent}, the positive class; $74$ \textsc{consistent}). Judge-based
rows show the median-F1 run of three repeats; the rule-based monitor is
deterministic. The best value in each column is \textbf{bold}.}
\label{tab:main-results}
\centering
\footnotesize
\setlength{\tabcolsep}{2.6pt}
\begin{tabular}{@{}ll cccc ccc@{}}
\toprule
Monitor & Judge
& TP & FP & FN & TN
& Prec. & Rec. & F1 \\
\midrule
Rule-based   & N/A
& 9 & 13 & 17 & 61
& 0.41 & 0.35 & 0.38 \\
\midrule
Ego-frame  & Qwen3.5-4B
& 22 & 28 & 4 & 46
& 0.44 & 0.85 & 0.58 \\
               & DeepSeek V4.1 Flash
& 19 & 16 & 7 & 58
& 0.54 & 0.73 & 0.62 \\
               & Opus 5.5
& 18 & 15 & 8 & 59
& 0.55 & 0.69 & 0.61 \\
               & Jev 1.14
& 21 & 28 & 5 & 46
& 0.43 & 0.81 & 0.56 \\
\midrule
Lane-aware & Qwen3.5-4B
& \textbf{23} & 16 & \textbf{3} & 58
& 0.59 & \textbf{0.88} & 0.71 \\
     & DeepSeek V4.1 Flash
& 21 & 7 & 5 & 67
& 0.75 & 0.81 & 0.78 \\
     & Opus 5.5
& 21 & \textbf{4} & 5 & \textbf{70}
& \textbf{0.84} & 0.81 & \textbf{0.82} \\
     & Jev 1.14
& \textbf{23} & 22 & \textbf{3} & 52
& 0.51 & \textbf{0.88} & 0.65 \\
\bottomrule
\end{tabular}
\end{table}

\paragraph{Metrics}
Treating \textsc{inconsistent} as the positive class, we report precision,
recall, and $F_1$. Against consistency labels, a flagged \textsc{consistent} pair is a \emph{false positive}; against safety labels, a flagged \textsc{safe} pair is a \emph{false alarm}. For RQ4, we cross-tabulate the human consistency and safety labels and evaluate inconsistency as a safety signal by measuring its recall and precision for unsafe trajectories on the reliable subset. When comparing with the safety-monitor baselines on that subset, \textsc{unsafe} is the positive class, each monitor's alarm is the positive prediction, and we report precision, recall, and $F_1$. On the challenge set, which has no \textsc{safe} cases, we report alarm counts only.

\paragraph{Runtime evaluation (RQ5)}
For latency, we time the $100$ reliable pairs for every configuration of Table~\ref{tab:main-results} except Opus 5.5, whose batch API reports no per-query latency. We measure the latency of rule-based monitor from the start of trajectory parsing to its verdict, excluding file I/O. For judge-based monitors, we measure from issuing a query to the completion of output generation, excluding evidence and prompt construction. The rule-based monitor runs on one core of an Intel Xeon Gold 6248 CPU; Qwen3.5-4B is served locally with vLLM on one NVIDIA A100-80GB GPU;  DeepSeek V4.1 
Flash, Opus 5.5, and Jev 1.14 are queried
through hosted endpoints (latency also includes network transfer and
provider-side queuing). We compare latency against the $10$\,Hz per-plan rate, at which the VLA is queried.

\subsection{Results}


\paragraph{Detection accuracy (RQ3)}
Lane-relative evidence improves consistency detection across all four judges
(Table~\ref{tab:main-results}). Every lane-aware configuration outperforms every
ego-frame configuration and the rule-based monitor on the reliable subset. With the lane-aware monitor, the
$4$B-parameter Qwen3.5-4B ($F_1=0.71$) exceeds the frontier Opus 5.5 with raw
waypoints ($F_1=0.61$), and the strongest configuration, the lane-aware monitor with Opus 5.5, reaches
$F_1=0.82$.

The improvement comes mainly from fewer false positives, many caused by confusing road
geometry with lateral maneuvers. Lane-relative evidence distinguishes following a
curved lane from moving across lanes (e.g., Fig.~\ref{fig:fllm-recovery}, where a
``nudge left'' CoT accompanies a trajectory that ends $6.58$\,m right of the ego's
initial heading but moves left relative to the routed lane). Across the four judges,
the lane-aware monitor corrects false positives on $21$ distinct pairs, and on $16$ of them the ego and
lane frames differ by at least $1$\,m in lateral displacement.
It also raises TP count by $1$ to $3$, catching trajectories that match the CoT only
in the ego frame.

\begin{figure}[t]
\centering
\definecolor{fgInk}{HTML}{23272E}
\definecolor{fgRule}{HTML}{C5CAD3}
\definecolor{fgOK}{HTML}{1B7837}
\definecolor{fgBad}{HTML}{B2182B}
\newcommand{\citedval}[1]{{\setlength{\fboxsep}{.8pt}\colorbox{fgBad!16}{\textup{\textbf{#1}}}}}
\begin{tikzpicture}[x=1cm,y=1cm,font=\scriptsize,text=fgInk,line cap=round,
  every node/.style={execute at begin node={\hyphenpenalty=10000\exhyphenpenalty=10000}},
  chip/.style={fill=white,rounded corners=1pt,inner sep=1.5pt,font=\scriptsize\bfseries,anchor=north west}]
\begin{scope}
  \clip[rounded corners=2pt] (0,0) rectangle (6.02,-2.93);
  \node[anchor=north west,inner sep=0] at (0,0) {\includegraphics[width=6.02cm,height=2.93cm]{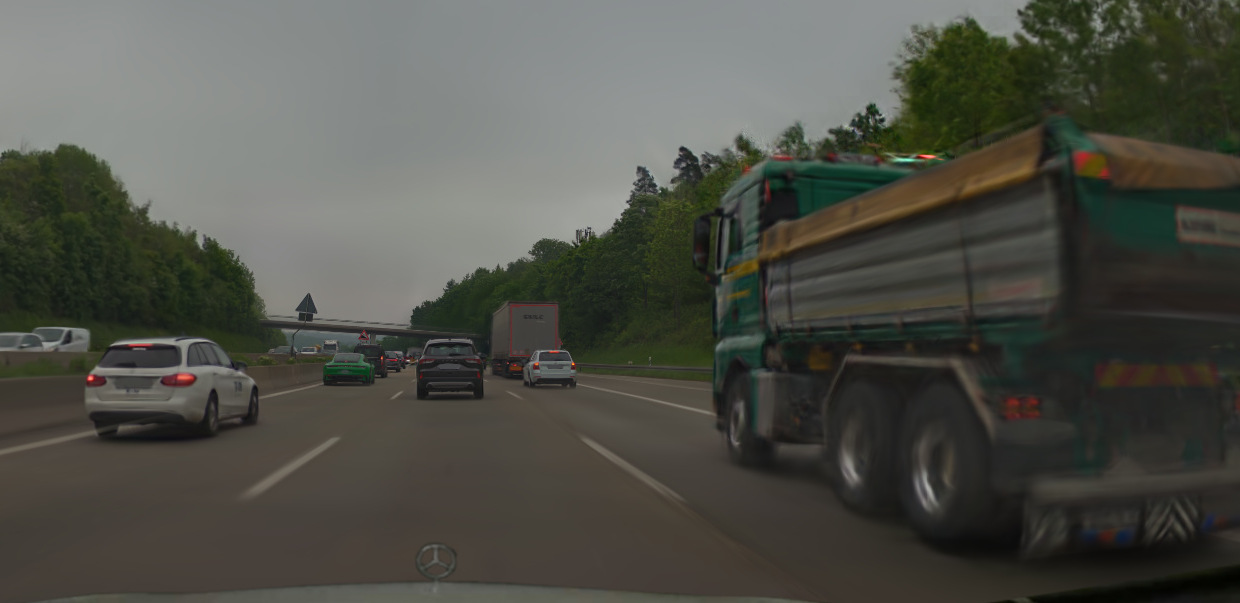}};
\end{scope}
\draw[white,line width=1pt] (2.0,-1.628) rectangle (2.37,-1.944);
\draw[fgBad,line width=.55pt] (2.0,-1.628) rectangle (2.37,-1.944);
\draw[white,line width=1.1pt,<->,>={Stealth[length=1.4mm,width=1.1mm]}] (2.185,-1.97) -- (2.185,-2.85);
\draw[fgBad,line width=.5pt,<->,>={Stealth[length=1.3mm,width=1mm]}] (2.185,-1.97) -- (2.185,-2.85);
\draw[fgRule,line width=.45pt,rounded corners=2pt] (0,0) rectangle (6.02,-2.93);
\node[chip] at (.07,-.07) {(a) Scene};
\node[anchor=north west,inner sep=0,font=\footnotesize\bfseries] at (6.2,0) {(b) Prompt evidence};
\node[anchor=north west,inner sep=0] at (6.2,-.33) {\setlength{\tabcolsep}{3.5pt}\renewcommand{\arraystretch}{0.9}%
  \begin{tabular}{@{}rrr@{}}
    \toprule
    $t$ (s) & $v$ (m/s) & $a$ (m/s$^2$) \\
    \midrule
    0.0 & \citedval{9.26} & 0.00 \\
    0.8 & 9.22 & $-$0.15 \\
    1.6 & 9.11 & $-$0.11 \\
    2.4 & 9.07 & 0.04 \\
    3.2 & 9.25 & 0.50 \\
    4.0 & 9.74 & 0.56 \\
    4.8 & 10.15 & 0.42 \\
    5.6 & 10.34 & $-$0.06 \\
    6.4 & \citedval{10.21} & $-$0.37 \\
    \bottomrule
  \end{tabular}};
\node[anchor=north west,inner sep=0] at (0,-3.04) {\textbf{CoT:} {\itshape ``Keep distance to the lead vehicle since it is moving slowly ahead.''}};
\node[anchor=north west,rounded corners=2pt,draw=fgBad,fill=fgBad!5,line width=.45pt,text width=8.03cm,inner xsep=.22cm,inner ysep=.09cm,align=left] (v) at (0,-3.36) {%
  Ego-frame \& lane-aware: $2/5$ \textcolor{fgBad}{\textsc{inconsistent}}\hfill\tikz[baseline=(c.base)]\node[fill=fgBad,rounded corners=1pt,inner sep=1.2pt,text=white,font=\scriptsize\bfseries] (c) {both wrong};\\[2pt]
  \textbf{Lane-aware rationale:} {\itshape ``\ldots implying longitudinal caution (holding or slowing to avoid closing the
  gap), but the trajectory shows net acceleration increasing speed from
  \citedval{9.26} to \citedval{10.21}~m/s, conflicting with the safety intent of
  keeping distance.''}};
\fill[fgBad,rounded corners=1pt] (v.north west) rectangle ([xshift=.07cm]v.south west);
\end{tikzpicture}
\caption{Shared error of the ego-frame and lane-aware monitors (judge:
Qwen3.5-4B). Annotators label the pair \textsc{consistent}, but both monitors score
it \textsc{inconsistent} ($2/5$), reading mild ego acceleration as gap closing. The
ego-frame monitor's rationale cites the same speeds plus $0.91$\,m of lateral drift.
``Keep distance'' is relational over the ego--lead gap, but the evidence contains
only ego motion, so agent-relative predicates require agent-relative evidence.}
\label{fig:fllm-error}
\end{figure}

The lane-aware monitor's remaining errors point to two limitations of the current
evidence representation and judging setup. First, judges can read the CoT too
literally. For example, Qwen3.5-4B rejects a ``stop'' CoT because a
stationary ego creeps at $0.02$--$0.06$\,m/s and never reaches exactly zero. It also
rejects a ``nudge'' as too large because the ego moves about $3$\,m sideways, but
annotators accept it because the ego does not change lanes, moving from $1.8$\,m right
of center to $1.4$\,m left.
Second, neither raw waypoints nor lane-relative evidence includes \emph{agent-relative
evidence}, which tracks other agents relative to the ego, such as the ego--lead gap, so
judges must infer ``keep distance'' from ego speed and
acceleration. Qwen3.5-4B
and DeepSeek V4.1 Flash can read mild acceleration as gap closing. In Fig.~\ref{fig:fllm-error},
for example, Qwen3.5-4B flags a ``keep distance'' pair because of a mild
acceleration, although annotators label the pair consistent and safe.

These results suggest that, for reliable CoTs, evidence representation
improves every judge, although it does not resolve literal readings of the
CoT or claims about other agents.

\paragraph{Safety relevance (RQ4)} 
Using human consistency labels, we first ask whether inconsistency itself is
safety-relevant. Table~\ref{tab:cons-safety} starts from
trajectory-safety annotations and reports the consistency labels within each
safety class, on the reliable subset and, separately, on the challenge set. Of the $9$ unsafe trajectories in the
reliable subset, $7$ are \textsc{inconsistent}, so inconsistency labels cover
$7/9=77.8\%$ of unsafe trajectories, against $19/91=20.9\%$ of safe
ones; Fig.~\ref{fig:teaser}(b)
illustrates one such case, where the ego's planned trajectory
diverges from its stated maneuver in an unsafe way. The challenge set
contains only unsafe pairs, so it has no safe row; $10/15=66.7\%$ of its pairs
are \textsc{inconsistent}, a similar share. Conversely, on the reliable subset (which includes both safe and unsafe trajectories), $7/26=26.9\%$ of \textsc{inconsistent} pairs are unsafe, higher than $2/74=2.7\%$ among \textsc{consistent} pairs, suggesting that inconsistency may serve as a safety signal.

\begin{table}[t]
\caption{Human consistency labels by trajectory safety. The challenge set
contains only unsafe pairs.}
\label{tab:cons-safety}
\centering
\small
\setlength{\tabcolsep}{7pt}
\begin{tabular}{@{}lrrrr@{}}
\toprule
Trajectory safety & \textsc{Incons.} & \textsc{Cons.} & Total & Incons.\ rate \\
\midrule
\multicolumn{5}{@{}l}{\emph{Reliable subset ($n=100$)}} \\
\textsc{Unsafe} & 7  & 2  & 9   & 77.8\% \\
\textsc{Safe}   & 19 & 72 & 91  & 20.9\% \\
Total           & 26 & 74 & 100 & 26.0\% \\
\midrule
\multicolumn{5}{@{}l}{\emph{Challenge set ($n=15$)}} \\
\textsc{Unsafe} & 10 & 5  & 15  & 66.7\% \\
\bottomrule
\end{tabular}
\end{table}

\begin{table*}[t]
\caption{Unsafe-trajectory detection on the reliable subset ($9$ \textsc{unsafe},
$91$ \textsc{safe}) and the challenge set ($15$ \textsc{unsafe}, no \textsc{safe}).
\emph{Unsafe flagged} counts \textsc{unsafe} cases that raise an alarm, and
\emph{False alarms} counts \textsc{safe} cases that do. On the reliable subset, judge-based
rows use the median-F1 runs of Table~\ref{tab:main-results}, chosen on
inconsistency-detection F1, and SelfOracle its median-F1 seed. Challenge-set
counts are medians over a monitor's repeats (seeds for SelfOracle). The best value in
each column is \textbf{bold}.}
\label{tab:safety-baselines}
\centering
\small
\setlength{\tabcolsep}{5pt}
\begin{tabular}{@{}ll cc ccc c@{}}
\toprule
& & \multicolumn{5}{c}{Reliable subset} & Challenge set \\
\cmidrule(lr){3-7}\cmidrule(l){8-8}
Monitor & Variant & Unsafe flagged (of $9$) & False alarms (of $91$) & Prec. & Rec. & F1 & Unsafe flagged (of $15$) \\
\midrule
\multicolumn{8}{@{}l}{\emph{Safety-monitor baselines}} \\
SelfOracle & SAE & 3 & 16 & 0.16 & 0.33 & 0.21 & 1 \\
           & VAE & 0 & \textbf{3} & 0.00 & 0.00 & 0.00 & 2 \\
\baselineB & $2$\,s & 4 & 33 & 0.11 & 0.44 & 0.17 & 4 \\
\midrule
\multicolumn{8}{@{}l}{\emph{Consistency monitors}} \\
Rule-based & N/A & 2 & 20 & 0.09 & 0.22 & 0.13 & 4 \\
Ego-frame & Qwen3.5-4B
& 6 & 44 & 0.12 & 0.67 & 0.20 & 9 \\
    & DeepSeek V4.1 Flash
& 5 & 30 & 0.14 & 0.56 & 0.23 & 8 \\
    & Opus 5.5
& 5 & 28 & 0.15 & 0.56 & 0.24 & 7 \\
    & Jev 1.14
& 6 & 43 & 0.12 & 0.67 & 0.21 & 8 \\
Lane-aware & Qwen3.5-4B
& \textbf{7} & 32 & 0.18 & \textbf{0.78} & 0.29 & 8 \\
      & DeepSeek V4.1 Flash
& \textbf{7} & 21 & 0.25 & \textbf{0.78} & 0.38 & 7 \\
      & Opus 5.5
& \textbf{7} & 18 & \textbf{0.28} & \textbf{0.78} & \textbf{0.41} & 7 \\
      & Jev 1.14
& \textbf{7} & 38 & 0.16 & \textbf{0.78} & 0.26 & \textbf{10} \\
\bottomrule
\end{tabular}
\end{table*}


\paragraph{Comparison with safety monitors}

The monitors detect inconsistency imperfectly (Table~\ref{tab:main-results}), so the
safety signal in the human labels may not carry over to their alarms. For the lane-aware monitor, the signal largely carries over (Table~\ref{tab:safety-baselines}). In the runs selected by median inconsistency-detection $F_1$ (Table~\ref{tab:main-results}), the monitor flags all
$7$ unsafe trajectories that humans label \textsc{inconsistent} with every judge; the
other $2$ of the $9$ are \textsc{consistent} with the CoT and therefore outside the
scope of a perfect consistency monitor. The ego-frame
configurations flag fewer unsafe trajectories ($5$--$6$ of the $9$) and raise more false alarms
($28$--$44$, against $18$--$38$ for the lane-aware ones). For reference, the safety-monitor baselines,
which do not read the CoT, flag $0$--$4$ of the $9$. Precision is low for every monitor,
including the human labels ($0.27$), because only $9\%$ of pairs are unsafe and many
inconsistencies are benign. Even so, these results suggest that a performant
consistency monitor can help identify safety risk.

\paragraph{Unsafe pairs in the challenge set}
The reliable subset contains only $9$ unsafe trajectories, so we also count alarms on
the $15$ challenge pairs, all of them unsafe. The judge-based monitors still flag
most of the \textsc{inconsistent} ones: at the median over repeats, $6$--$9$ of the
$10$, depending on the judge, and $7$--$10$ of all $15$
(Table~\ref{tab:safety-baselines}). Contrary to results in the reliable subset, these
counts do not clearly show whether the ego-frame or lane-aware monitor is better. The $5$
\textsc{consistent} unsafe plans lie outside what a consistency monitor can detect, so
its $0$--$2$ alarms on them are in the table but are not detections. For reference,
the rule-based monitor flags $4$ of the $15$ ($3$ of the $10$), SelfOracle $1$ (SAE) and
$2$ (VAE), and the \baselineB\ monitor $4$.

Together, these results suggest that, for reliable CoTs, inconsistency is a useful
safety signal. Most unsafe trajectories in both sets are inconsistent,
and the judge-based monitors flag most of those, but some unsafe
trajectories are consistent and false alarms are common on the reliable subset.

\begin{table}[t]
\caption{Approximate per-pair latency on the $100$ reliable pairs across three repeats (per-plan rate: $100$\,ms / $10$\,Hz). \textbf{Med./P95}: median and 95th-percentile latency (s); \textbf{Tok.}: mean generated tokens (including thinking); \textbf{Rate}: sequential verdict rate ($1/\text{Med.}$, in Hz). \textnormal{`--'}: inapplicable (Opus~5.5 ran via batch API; Rule-based and Jev 1.14 generate no text tokens).}
\label{tab:latency}
\centering
\footnotesize
\setlength{\tabcolsep}{3.5pt}
\begin{tabular}{@{}ll rrrr@{}}
\toprule
Monitor & Judge & Med. & P95 & Tok. & Rate \\
\midrule
Rule-based & N/A
& 0.0021 & 0.0025 & -- & 476 \\
\midrule
Ego-frame & Qwen3.5-4B
& 27.2 & 45.0 & 4{,}395 & 0.037 \\
    & DeepSeek V4.1 Flash
& 2.91 & 12.7 & 657 & 0.34 \\
    & Opus 5.5
& -- & -- & 395 & -- \\
    & Jev 1.14
& 0.093 & 0.161 & -- & 10.8 \\
\midrule
Lane-aware & Qwen3.5-4B
& 30.5 & 45.6 & 4{,}757 & 0.033 \\
      & DeepSeek V4.1 Flash
& 3.75 & 12.3 & 691 & 0.27 \\
      & Opus 5.5
& -- & -- & 391 & -- \\
      & Jev 1.14
& 0.102 & 0.156 & -- & 9.8 \\
\bottomrule
\end{tabular}
\end{table}

\paragraph{Inference latency (RQ5)}
Table~\ref{tab:latency} reports approximate per-pair latency. The rule-based monitor runs far above the per-plan rate on one CPU core ($2.1$\,ms median, $476$\,Hz), although its $F_1$ of $0.38$ is the lowest in Table~\ref{tab:main-results}. With the same judge, the lane-aware monitor's verdict rate is lower than the ego-frame monitor's. Used
off-the-shelf in thinking mode, in which a judge generates hidden tokens before
its answer, every generative judge we timed is too slow for per-plan monitoring. With the lane-aware monitor, the hosted DeepSeek V4.1 Flash takes $3.75$\,s at the median
and generates $691$ tokens, mostly thinking, so a single stream checks roughly
one in every $38$ plans. The local Qwen3.5-4B judge is about eight times
slower ($30.5$\,s median, roughly one in every $305$ plans) and generates about
seven times as many tokens ($4{,}757$). However, Jev 1.14 achieves $>10$\,Hz with the ego-frame monitor and nearly $10$\,Hz with the lane-aware monitor (even over a hosted endpoint). This suggests that
autoregressive thinking and rationale generation are the primary latency
bottleneck for generative judges, making decision models a
promising direction for per-plan monitoring. Per-plan checking may also be unnecessary: because consecutive plans overlap over a multi-second horizon, runtime operation may not require checking every planning step. We leave the required checking frequency to future work.

\subsection{Threats to Validity}
\label{subsec:threats}
\noindent\textbf{Internal validity.} Our results depend on how we configured the
judge-based monitors: the judge prompt and scoring rubric, the threshold $\theta$,
and, for the lane-aware monitor, the selected route and the lane-relative evidence computed from
it. That monitor also adapts the rubric and few-shot examples to this evidence, so its
$F_1$ gains reflect the two together. Two other monitors are our own versions of prior work. The rule-based monitor's thresholds, phrase dictionary,
and matcher reconstruct Alpamayo-R1's reasoning--action consistency procedure from the
available description~\cite{alpamayo-r1} and may differ from the original. Likewise, SelfOracle is our adaptation of the released code to AlpaSim camera streams, with our own nominal training data
and preprocessing, and it may perform differently in its original setting.

\noindent\textbf{Construct validity.} Human annotations are our reference for
CoT reliability, CoT--trajectory consistency, and trajectory safety, but the benchmark is not fully
double-annotated. One annotator labeled most pairs, and a second independently
labeled only $10$ pairs. Agreement on these double-annotated pairs was high ($9/10$ for reliability and $10/10$ for consistency and safety), but it was lower on the challenge set, so high agreement may not hold across
the full benchmark. We also evaluate the monitors offline on recorded pairs, not with alarms
acting on the driving, so our results do not show whether acting on an alarm would make
driving safer.

\noindent\textbf{External validity.} Our benchmark evaluates a single reasoning
VLA, Alpamayo~1.5, which was the state of the art when we collected and annotated
the data. Newer models have since been released, including its successor,
Alpamayo~2 Super~\cite{nvidia_alpamayo2_model_card}, which became runnable in
AlpaSim only after our initial data collection. These models could exhibit
different reasoning behaviors and capabilities, so our findings may not fully
generalize to them. We also evaluate in a single simulator, AlpaSim, because it runs Alpamayo in its released configuration and reconstructs scenes from real camera images, keeping the VLA’s inputs close to real driving. However, it does not expose
some information, such as traffic light states and physical boundaries, that
safety monitors such as ARGUS~\cite{wang_argus_2025} and scene-flow
monitors~\cite{woodlief_sceneflow_2025} need to work properly. Our safety-monitor
baselines, SelfOracle and \baselineB, are therefore simple reference points
rather than state-of-the-art safety monitors, and stronger CoT-blind monitors may
detect more unsafe trajectories than reported here. Finally, we sampled benchmark
pairs for diversity rather than from a naturalistic driving distribution. Considering the threats, our
reported rates should therefore be read as conditional on this setup rather than
as population estimates for deployed AVs.

\section{Discussion}
\label{sec:discussion}
\paragraph{Score threshold}
All experiments flag a pair as \textsc{inconsistent} when the judge's score is at
most $\theta=2$. Varying $\theta$ from $1$ to $4$ on the same runs
(Appendix~\ref{app:threshold}) trades precision for recall, and the default gives the
highest $F_1$ (tied to two decimal places) in all eight configurations. At $\theta=1$, the lane-aware monitor's alarms
are $76.9$--$87.5\%$ precise for every judge but catch only $26.9$--$42.3\%$ of
inconsistencies, which favors a costly response such as a fallback maneuver. At
$\theta=4$, every judge-based configuration recovers $88.5$--$100\%$ of inconsistencies, but precision
falls to $30.5$--$45.5\%$, which suits only a cheap response, such as logging the plan
for review. The lane-aware monitor has the higher $F_1$ at every $\theta\le3$ for every judge.

\paragraph{Temporal aggregation}
A driving VLA replans frequently (e.g., every $100$\,ms), so an inconsistent plan may
be replaced by a corrected one before much of it is executed. Raising an alarm only
when inconsistency persists over consecutive plans could therefore reduce false
positives. We test this on the $92$ reliable pairs with enough planning history, for
the lane-aware monitor with Qwen3.5-4B. On those 92 pairs, requiring two consecutive flags raises precision from $0.58$ to
$0.88$ but lowers recall from $0.88$ to $0.60$, leaving $F_1$ almost unchanged ($0.70$
to $0.71$). Appendix~\ref{app:temporal} gives the setup and other alarm rules.
\section{Conclusion}

This paper studies whether CoTs from driving VLAs can
serve as specifications for runtime monitoring. We identify two requirements for this
idea to be meaningful: the CoT must first be reliable, and the generated trajectory must then be comparable to it. On
DriveAlignBench, a 150-pair Alpamayo~1.5 benchmark, $100/150$ CoTs satisfy our grounding, safety,
and coherence criteria. Reliability is strongest for routine, directly
observable behavior such as lead following, lane keeping, and set-speed tracking,
and weaker for decisions that require inferring lane topology, routing intent, or
spatial and temporal gaps. These results suggest that CoTs can provide useful
runtime specifications, but only selectively and with explicit reliability
filtering.

We then develop rule-based, ego-frame, and lane-aware monitors
for checking whether a trajectory realizes a reliable CoT. On the reliable
subset, the best lane-aware configuration reaches $F_1=0.82$, against $0.38$ for the
rule-based monitor. Lane-relative evidence raises $F_1$ for all four
judges, from $0.56$--$0.62$ with raw waypoints to $0.65$--$0.82$, suggesting that
evidence representation can improve detection across judges. For reliable CoTs,
inconsistency is also associated with unsafe behavior: $27\%$ of
human-labeled \textsc{inconsistent} trajectories are unsafe, against $3\%$ of
\textsc{consistent} ones. Most unsafe trajectories are \textsc{inconsistent} ($7/9$ on the
reliable subset and $10/15$ on a separate challenge set from hazardous scenes), and the lane-aware monitor flags most of
those, but some unsafe trajectories are \textsc{consistent} and false alarms are common,
so inconsistency alone is not a complete safety monitor. The remaining failures point to broader limits of judge-based monitoring.
Judges can over-weight surface action words and lack the agent-relative evidence
needed to check claims such as ``keep distance'' from ego motion alone. Future work should first close three gaps in the monitoring pipeline.
CoT reliability should be screened at runtime, because our monitors assume a reliable CoT
and we identify those with human labels. Monitors should move
beyond lane geometry to agent-relative evidence tracked over time, so that claims such as
``yield'' or ``merge behind'' can be checked. Judges should reach the per-plan rate
without losing accuracy. Evaluations should then extend to additional VLAs and real-world
data, and monitor-in-the-loop studies should test whether these monitors can trigger fallback or
replanning before inconsistent trajectories become hazards.

\bibliographystyle{ACM-Reference-Format}
\bibliography{references}

@misc{alpamayo-r1,
  title         = {{Alpamayo-R1}: Bridging Reasoning and Action Prediction for Generalizable Autonomous Driving in the Long Tail},
  author        = {{NVIDIA} and Wang, Yan and Luo, Wenjie and Bai, Junjie and Cao, Yulong and Che, Tong and Chen, Ke and Chen, Yuxiao and Diamond, Jenna and Ding, Yifan and Ding, Wenhao and Feng, Liang and Heinrich, Greg and Huang, Jack and Karkus, Peter and Li, Boyi and Li, Pinyi and Lin, Tsung-Yi and Liu, Dongran and Liu, Ming-Yu and Liu, Langechuan and Liu, Zhijian and Lu, Jason and Mao, Yunxiang and Molchanov, Pavlo and Pavao, Lindsey and Peng, Zhenghao and Ranzinger, Mike and Schmerling, Ed and Shen, Shida and Shi, Yunfei and Tariq, Sarah and Tian, Ran and Wekel, Tilman and Weng, Xinshuo and Xiao, Tianjun and Yang, Eric and Yang, Xiaodong and You, Yurong and Zeng, Xiaohui and Zhang, Wenyuan and Ivanovic, Boris and Pavone, Marco},
  year          = {2025},
  eprint        = {2511.00088},
  archivePrefix = {arXiv}
}

@misc{nvidia_alpamayo15_model_card,
  title        = {{Alpamayo} 1.5 Model Card},
  author       = {{NVIDIA}},
  year         = {2026},
  howpublished = {Hugging Face model repository},
  note         = {Release date: March 19, 2026. Accessed: 2026-07-01},
  url          = {https://huggingface.co/nvidia/Alpamayo-1.5-10B}
}

@misc{nvidia_alpamayo2_model_card,
  title        = {{Alpamayo} 2 Super Model Card},
  author       = {{NVIDIA}},
  year         = {2026},
  howpublished = {Hugging Face model repository},
  note         = {Release date: August 4, 2026. Accessed: 2026-11-04},
  url          = {https://huggingface.co/nvidia/Alpamayo2-Super}
}

@misc{nvidia_alpasim,
  author = {{NVIDIA} and Cao, Yulong and de Lutio, Riccardo and Fidler, Sanja and
            Garcia Cobo, Guillermo and Gojcic, Zan and Igl, Maximilian and
            Ivanovic, Boris and Karkus, Peter and Martinez Esturo, Janick and
            Pavone, Marco and Smith, Aaron and Tanimura, Ellie and
            Tyszkiewicz, Michal and Watson, Michael and Wu, Qi and Zhang, Le},
  title  = {{AlpaSim}: A Modular, Lightweight, and Data-Driven Research Simulator for Autonomous Driving},
  year   = {2025},
  month  = oct,
  url    = {https://github.com/NVlabs/alpasim}
}

@inproceedings{li2022hdmapnet,
  title     = {{HDMapNet}: An Online {HD} Map Construction and Evaluation Framework},
  author    = {Li, Qi and Wang, Yue and Wang, Yilun and Zhao, Hang},
  booktitle = {ICRA},
  pages     = {4628--4634},
  year      = {2022},
  doi       = {10.1109/ICRA46639.2022.9812383},
  url       = {https://arxiv.org/abs/2107.06307}
}

@misc{openai_gpt55,
  title        = {Introducing {GPT-5.5}},
  author       = {{OpenAI}},
  year         = {2026},
  month        = apr,
  howpublished = {OpenAI product release},
  note         = {Published April 23, 2026. Accessed: 2026-07-01},
  url          = {https://openai.com/index/introducing-gpt-5-5/}
}

@misc{anthropic_claude_opus55,
  title        = {Introducing {Claude Opus 5.5}},
  author       = {{Anthropic}},
  year         = {2026},
  month        = sep,
  howpublished = {Anthropic product release},
  note         = {Published September 22, 2026. Accessed: 2026-10-31},
  url          = {https://www.anthropic.com/claude-opus-5-5}
}

@misc{redhat_qwen35_4b_fp8,
  title        = {{Qwen3.5-4B-FP8-dynamic} Model Card},
  author       = {{Red Hat AI}},
  year         = {2026},
  howpublished = {Hugging Face model repository},
  note         = {Accessed: 2026-07-01},
  url          = {https://huggingface.co/RedHatAI/Qwen3.5-4B-FP8-dynamic}
}

@techreport{ntsb_uber_tempe_2019,
  author      = {{National Transportation Safety Board}},
  title       = {Collision Between Vehicle Controlled by Developmental Automated Driving System and Pedestrian, Tempe, Arizona, March 18, 2018},
  institution = {National Transportation Safety Board},
  type        = {Highway Accident Report},
  number      = {NTSB/HAR-19/03},
  address     = {Washington, DC},
  year        = {2019},
  note        = {PB2019-101402},
  url         = {https://www.ntsb.gov/investigations/AccidentReports/Reports/HAR1903.pdf}
}

@techreport{ntsb_tesla_mountain_view_2020,
  author      = {{National Transportation Safety Board}},
  title       = {Collision Between a Sport Utility Vehicle Operating With Partial Driving Automation and a Crash Attenuator, Mountain View, California, March 23, 2018},
  institution = {National Transportation Safety Board},
  type        = {Highway Accident Report},
  number      = {NTSB/HAR-20/01},
  address     = {Washington, DC},
  year        = {2020},
  note        = {PB2020-100112},
  url         = {https://www.ntsb.gov/investigations/AccidentReports/Reports/HAR2001.pdf}
}

@misc{nhtsa_cruise_consent_order_2024,
  author       = {{National Highway Traffic Safety Administration}},
  title        = {{NHTSA} Announces Consent Order with Cruise After Company Failed to Fully Report Crash Involving Pedestrian},
  howpublished = {Press release},
  year         = {2024},
  month        = sep,
  note         = {September 30, 2024},
  url          = {https://www.nhtsa.gov/press-releases/consent-order-cruise-crash-reporting}
}

@article{chen_e2e_ad_survey_2024,
  author  = {Chen, Li and Wu, Penghao and Chitta, Kashyap and Jaeger, Bernhard and Geiger, Andreas and Li, Hongyang},
  title   = {End-to-End Autonomous Driving: Challenges and Frontiers},
  journal = {IEEE TPAMI},
  volume  = {46},
  number  = {12},
  pages   = {10164--10183},
  year    = {2024},
  doi     = {10.1109/TPAMI.2024.3435937},
  url     = {https://arxiv.org/abs/2306.16927}
}

@misc{wayve_lingo2_2024,
  author       = {{Wayve}},
  title        = {{LINGO-2}: Driving with Natural Language},
  howpublished = {Wayve blog},
  year         = {2024},
  month        = apr,
  note         = {Published April 17, 2024},
  url          = {https://wayve.ai/thinking/lingo-2-driving-with-language/}
}

@article{hwang_emma_2025,
  author        = {Hwang, Jyh-Jing and Xu, Runsheng and Lin, Hubert and Hung, Wei-Chih and Ji, Jingwei and Choi, Kristy and Huang, Di and He, Tong and Covington, Paul and Sapp, Benjamin and Zhou, Yin and Guo, James and Anguelov, Dragomir and Tan, Mingxing},
  title         = {{EMMA}: End-to-End Multimodal Model for Autonomous Driving},
  journal       = {TMLR},
  year          = {2025},
  eprint        = {2410.23262},
  archivePrefix = {arXiv}
}

@inproceedings{toledo_sgsm_2024,
  author    = {Toledo, Felipe and Woodlief, Trey and Elbaum, Sebastian and Dwyer, Matthew B.},
  title     = {Specifying and Monitoring Safe Driving Properties with Scene Graphs},
  booktitle = {ICRA},
  pages     = {15577--15584},
  year      = {2024},
  doi       = {10.1109/ICRA57147.2024.10610973}
}

@article{woodlief_sceneflow_2025,
  author  = {Woodlief, Trey and Toledo, Felipe and Dwyer, Matthew B. and Elbaum, Sebastian},
  title   = {Scene Flow Specifications: Encoding and Monitoring Rich Temporal Safety Properties of Autonomous Systems},
  journal = {Proc. ACM Softw. Eng.},
  volume  = {2},
  number  = {FSE},
  pages   = {2524--2547},
  year    = {2025},
  doi     = {10.1145/3729382},
  url     = {https://doi.org/10.1145/3729382}
}

@inproceedings{wei_cot_2022,
  author        = {Wei, Jason and Wang, Xuezhi and Schuurmans, Dale and Bosma, Maarten and Ichter, Brian and Xia, Fei and Chi, Ed and Le, Quoc V. and Zhou, Denny},
  title         = {Chain-of-Thought Prompting Elicits Reasoning in Large Language Models},
  booktitle     = {NeurIPS},
  volume        = {35},
  pages         = {24824--24837},
  year          = {2022},
  doi           = {10.52202/068431-1800},
  url           = {https://arxiv.org/abs/2201.11903}
}

@misc{drivecot_2024,
  author        = {Wang, Tianqi and Xie, Enze and Chu, Ruihang and Li, Zhenguo and Luo, Ping},
  title         = {{DriveCoT}: Integrating Chain-of-Thought Reasoning with End-to-End Driving},
  year          = {2024},
  eprint        = {2403.16996},
  archivePrefix = {arXiv}
}

@inproceedings{tian_drivevlm_2024,
  author        = {Tian, Xiaoyu and Gu, Junru and Li, Bailin and Liu, Yicheng and Wang, Yang and Zhao, Zhiyong and Zhan, Kun and Jia, Peng and Lang, Xianpeng and Zhao, Hang},
  title         = {{DriveVLM}: The Convergence of Autonomous Driving and Large Vision-Language Models},
  booktitle     = {CoRL},
  year          = {2025},
  series        = {PMLR},
  volume        = {270},
  pages         = {4698--4726},
  eprint        = {2402.12289},
  archivePrefix = {arXiv}
}

@inproceedings{renz_simlingo_2025,
  author        = {Renz, Katrin and Chen, Long and Arani, Elahe and Sinavski, Oleg},
  title         = {{SimLingo}: Vision-Only Closed-Loop Autonomous Driving with Language-Action Alignment},
  booktitle     = {CVPR},
  year          = {2025},
  pages         = {11993--12003},
  doi           = {10.1109/CVPR52734.2025.01120},
  url           = {https://arxiv.org/abs/2503.09594}
}

@misc{mayumu_2026,
  author        = {Mayumu, Nicanor and Deng, Xiaoheng and Mukala, Patrick},
  title         = {Is {VLA} Reasoning Faithful? Probing Safety of Chain-of-Causation in Autonomous Driving Models},
  year          = {2026},
  eprint        = {2605.17268},
  archivePrefix = {arXiv}
}

@inproceedings{wu_2025,
  author        = {Wu, Yilin and Li, Anqi and Hermans, Tucker and Ramos, Fabio and Bajcsy, Andrea and P{\'e}rez-D'Arpino, Claudia},
  title         = {Do What You Say: Steering Vision-Language-Action Models via Runtime Reasoning-Action Alignment Verification},
  booktitle     = {ICRA},
  year          = {2026},
  pages         = {21113--21120},
  doi           = {10.1109/ICRA57385.2026.11696392},
  url           = {https://arxiv.org/abs/2510.16281}
}

@inproceedings{turpin_2023,
  author        = {Turpin, Miles and Michael, Julian and Perez, Ethan and Bowman, Samuel R.},
  title         = {Language Models Don't Always Say What They Think: Unfaithful Explanations in Chain-of-Thought Prompting},
  booktitle     = {NeurIPS},
  volume        = {36},
  year          = {2023},
  pages         = {74952--74965},
  doi           = {10.52202/075280-3275},
  url           = {https://arxiv.org/abs/2305.04388}
}

@misc{lanham_2023,
  author        = {Lanham, Tamera and Chen, Anna and Radhakrishnan, Ansh and Steiner, Benoit and Denison, Carson and others},
  title         = {Measuring Faithfulness in Chain-of-Thought Reasoning},
  year          = {2023},
  eprint        = {2307.13702},
  archivePrefix = {arXiv}
}

@inproceedings{luo_adathinkdrive_2025,
  author        = {Luo, Yuechen and Li, Fang and Xu, Shaoqing and Lai, Zhiyi and Yang, Lei and Chen, Qimao and Luo, Ziang and Xie, Zixun and Jiang, Shengyin and Liu, Jiaxin and Chen, Long and Wang, Bing and Yang, Zhi-xin},
  title         = {{AdaThinkDrive}: Adaptive Thinking via Reinforcement Learning for Autonomous Driving},
  booktitle     = {ICRA},
  year          = {2026},
  pages         = {14489--14496},
  doi           = {10.1109/ICRA57385.2026.11696257},
  url           = {https://arxiv.org/abs/2509.13769}
}

@inproceedings{yuan_autodriver2_2026,
  author        = {Yuan, Zhenlong and Qian, Chengxuan and Tang, Jing and Chen, Rui and Song, Zijian and Sun, Lei and Chu, Xiangxiang and Cai, Yujun and Zhang, Dapeng and Li, Shuo},
  title         = {{AutoDrive-R}$^2$: Incentivizing Reasoning and Self-Reflection Capacity for {VLA} Model in Autonomous Driving},
  booktitle     = {ICLR},
  year          = {2026},
  eprint        = {2509.01944},
  archivePrefix = {arXiv}
}

@inproceedings{hendrycks_ood_2017,
  author        = {Hendrycks, Dan and Gimpel, Kevin},
  title         = {A Baseline for Detecting Misclassified and Out-of-Distribution Examples in Neural Networks},
  booktitle     = {ICLR},
  year          = {2017},
  eprint        = {1610.02136},
  archivePrefix = {arXiv}
}

@misc{shalev_rss_2017,
  author        = {Shalev-Shwartz, Shai and Shammah, Shaked and Shashua, Amnon},
  title         = {On a Formal Model of Safe and Scalable Self-Driving Cars},
  year          = {2017},
  eprint        = {1708.06374},
  archivePrefix = {arXiv}
}

@inproceedings{stocco_selforacle_2020,
  author    = {Stocco, Andrea and Weiss, Michael and Calzana, Marco and Tonella, Paolo},
  title     = {Misbehaviour Prediction for Autonomous Driving Systems},
  booktitle = {ICSE},
  pages     = {359--371},
  year      = {2020},
  doi       = {10.1145/3377811.3380353}
}

@inproceedings{hildebrandt_physcov_2023,
  author    = {Hildebrandt, Carl and von Stein, Meriel and Elbaum, Sebastian},
  title     = {{PhysCov}: Physical Test Coverage for Autonomous Vehicles},
  booktitle = {ISSTA},
  pages     = {449--461},
  year      = {2023},
  doi       = {10.1145/3597926.3598069}
}

@inproceedings{kuhn_introspective_2020,
  author    = {Kuhn, Christopher B. and Hofbauer, Markus and Petrovic, Goran and Steinbach, Eckehard},
  title     = {Introspective Black Box Failure Prediction for Autonomous Driving},
  booktitle = {IEEE IV},
  pages     = {1907--1913},
  year      = {2020},
  doi       = {10.1109/IV47402.2020.9304844}
}

@inproceedings{tang_actiongrounded_survey_2026,
  author        = {Tang, Zhengxu and Zhang, Xiaozhou and Cui, Guofeng and Gong, Ziyu and Wang, Zi and Shi, Yunfei and Deng, Ruifeng and Qi, Chengzhi and Chen, Ke and Patil, Sachin and Xiao, Tianjun and Liu, Langechuan and Wang, Pichao},
  title         = {Beyond Textual Chain-of-Thought: A Survey on Action-Grounded Reasoning in Autonomous Driving},
  booktitle     = {EMNLP Findings},
  year          = {2026},
  pages         = {11519--11561},
  doi           = {10.18653/2026.findings-emnlp.547},
  url           = {https://arxiv.org/abs/2609.01659}
}

@misc{jiang_workdrive_2026,
  author        = {Jiang, Tianyi and Zhang, Wen and Yang, Sihan and Lu, Ming and Zhang, Wentao},
  title         = {{WorkDrive}: Roadwork Chain of Causation for Autonomous Driving},
  year          = {2026},
  eprint        = {2607.14727},
  archivePrefix = {arXiv}
}

@misc{yang_dualprocess_2026,
  author        = {Yang, Zhongyao and Li, Haoyu and Yan, Yu and Yu, Zhuangxuan and Nan, Jiangfeng and Nan, Jinrui},
  title         = {Cognitive Dual-Process Planning for Autonomous Driving with Structured Scene Knowledge and Verifiable Reasoning-Action Consistency},
  year          = {2026},
  eprint        = {2607.19194},
  archivePrefix = {arXiv}
}

@misc{duzan_implicit_2026,
  author        = {Duzan, Agatha and Stickland, Asa Cooper},
  title         = {Chain-of-Thought Monitoring Can Be Unreliable in Implicit-Influence Settings},
  year          = {2026},
  eprint        = {2608.04735},
  archivePrefix = {arXiv}
}

@misc{severi_cotpoisoning_2026,
  author        = {Severi, Giorgio and Mirza, Shujaat and Bullwinkel, Blake and Minnich, Amanda},
  title         = {Evading Chain-of-Thought Monitoring Through Model Poisoning},
  year          = {2026},
  eprint        = {2608.02820},
  archivePrefix = {arXiv}
}

@misc{foutter_faithfulness_2026,
  author        = {Foutter, Matthew and Cercola, Matteo and Wild, Lena and Wang, Yunshan and Li, Michelle and Gammelli, Daniele and Pavone, Marco},
  title         = {Do Vision-Language-Action Models Mean What They Say? On the Role of Faithfulness in Embodied Reasoning},
  year          = {2026},
  eprint        = {2607.04681},
  archivePrefix = {arXiv}
}

@misc{nguyen_vladrivebench_2026,
  author        = {Nguyen, Thach and Guo, Danhua and Lampo, Tom and Wu, Fei and Yaman, Burhan},
  title         = {{VLADriveBench}: Evaluating {CoT}-Action Relationship in {VLA} for Autonomous Driving},
  year          = {2026},
  eprint        = {2606.12706},
  archivePrefix = {arXiv}
}

@misc{priyadershi_lostinfog_2026,
  author        = {Priyadershi, Abhinaw and Frtunikj, Jelena},
  title         = {Lost in Fog: Sensor Perturbations Expose Reasoning Fragility in Driving {VLAs}},
  year          = {2026},
  eprint        = {2605.21446},
  archivePrefix = {arXiv}
}

@inproceedings{park_hideandseek_2026,
  author        = {Park, Seongheon and Li, Wendi and Oh, Changdae and Yeh, Samuel and Kira, Zsolt and Hagenow, Michael and Li, Sharon},
  title         = {Hide-and-Seek in Trajectories: Discovering Failure Signals for {VLA} Runtime Monitoring},
  booktitle     = {NeurIPS},
  year          = {2026},
  eprint        = {2605.30834},
  archivePrefix = {arXiv}
}

@inproceedings{seligmann_vlafail_2026,
  author        = {Seligmann, Florian and Gospodinov, Emiliyan and Dincer, Enes Ulas and Neumann, Gerhard},
  title         = {{VLA-FAIL}: Efficient Task Failure Detection for Finetuned Vision-Language-Action Models},
  booktitle     = {CoRL},
  year          = {2026},
  eprint        = {2606.21386},
  archivePrefix = {arXiv}
}

@misc{bhardwaj_progressprobes_2026,
  author        = {Bhardwaj, Atiksh and Duan, Edward W. and Seo, Junwon and Dan, Prithwish and Ma, Wei-Chiu and Bajcsy, Andrea and Culbertson, Preston},
  title         = {Progress Probes {$P^2$}: Lightweight Runtime Monitoring for Vision-Language-Action Models},
  year          = {2026},
  eprint        = {2610.31613},
  archivePrefix = {arXiv}
}

@inproceedings{wang_argus_2025,
  author        = {Wang, Dingji and Lu, You and Chen, Bihuan and Hao, Shuo and Jiang, Haowen and Tian, Yifan and Peng, Xin},
  title         = {{Argus}: Resilience-Oriented Safety Assurance Framework for End-to-End {ADSs}},
  booktitle     = {ASE},
  year          = {2025},
  pages         = {13--25},
  doi           = {10.1109/ASE63991.2025.00010},
  url           = {https://arxiv.org/abs/2511.09032}
}

@misc{deepseek_v41_flash,
  title        = {{DeepSeek-V4.1-Flash-1017} Model Card},
  author       = {{DeepSeek-AI}},
  year         = {2026},
  month        = oct,
  howpublished = {Hugging Face model repository},
  note         = {Released October 17, 2026. Accessed: 2026-10-31},
  url          = {https://huggingface.co/deepseek-ai/DeepSeek-V4.1-Flash-1017}
}

@misc{typesafe_jev,
  title        = {{Jev 1.14} (jev-1.14-20261014)},
  author       = {{TypeSafe AI}},
  year         = {2026},
  month        = oct,
  howpublished = {Model page, OpenRouter},
  note         = {Released October 14, 2026. Accessed: 2026-11-01},
  url          = {https://openrouter.ai/typesafe/jev-1.14-20261014}
}

@misc{nvidia_physicalai_av_dataset,
  title        = {{PhysicalAI-Autonomous-Vehicles}},
  author       = {{NVIDIA}},
  year         = {2025},
  howpublished = {Dataset, Hugging Face},
  note         = {Accessed: 2026-11-03},
  url          = {https://huggingface.co/datasets/nvidia/PhysicalAI-Autonomous-Vehicles}
}

@misc{nvidia_cascade_dataset,
  title        = {{CASCADE}: Causal Spatio-Temporal Analysis of Driving Environments},
  author       = {{NVIDIA}},
  year         = {2026},
  howpublished = {Dataset, Hugging Face},
  note         = {Accessed: 2026-11-03},
  url          = {https://huggingface.co/datasets/nvidia/cascade}
}

@misc{nvidia_nurec_dataset,
  title        = {{PhysicalAI-Autonomous-Vehicles-NuRec}},
  author       = {{NVIDIA}},
  year         = {2025},
  howpublished = {Dataset, Hugging Face},
  note         = {Accessed: 2026-11-03},
  url          = {https://huggingface.co/datasets/nvidia/PhysicalAI-Autonomous-Vehicles-NuRec}
}

\clearpage
\appendix
\section*{Appendices}

\section{Safety-Monitor Baseline Settings}
\label{app:baselines}
\noindent\textbf{SelfOracle.} We train each autoencoder across three seeds on $630$
nominal ground-truth (GT) replay scenes ($125{,}969$ images) disjoint from the
benchmark. The alarm threshold is set to the $1-\epsilon$ quantile ($\epsilon=0.05$)
of a Gamma distribution fit to errors on $100$ held-out nominal scenes ($19{,}990$
images), averaging scores over the five frames ($0.4$\,s) ending at the pair's planning step. We
use all GT replay frames as nominal training data without anomaly filtering because
AlpaSim reconstructs 3D scenes directly from logged GT sensor data, yielding clean
renderings along the logged path. When Alpamayo deviates from this path during
rollout, rendering from poorly covered novel viewpoints introduces visual artifacts
that inflate reconstruction error, enabling SelfOracle to detect off-trajectory or
unsafe behavior.

\noindent\textbf{\baselineB.} Each time the VLA returns a trajectory, the monitor
constructs a trapezoidal region that the ego vehicle could reach within the next
$2$\,s. The region starts at the front bumper, is $2$\,s times the current ego speed
long, and widens by up to $\tfrac{1}{2}a_{\mathrm{lat}}(2\,\mathrm{s})^2 = 6$\,m on
each side for a lateral acceleration $a_{\mathrm{lat}} = 3$\,m/s$^2$. It raises an
alarm if any other road user's bounding box overlaps this region.

\section{Temporal Aggregation}
\label{app:temporal}
Because a driving VLA replans frequently (e.g., every $100$\,ms), an inconsistent plan may be superseded by a corrected one before much of it is executed. Aggregating monitor verdicts over consecutive planning steps may help identify persistent inconsistencies, which in hypothesis may correlate more strongly with safety-critical risks.
We treat a deployed monitor as emitting a stream of
verdicts $\hat{y}_t\in\{0,1\}$, with $1$ denoting \textsc{inconsistent}, from which an
alarm rule $\operatorname{Alarm}_t$ decides when to act. For each reliable target pair at
planning step $t$, we take the CoT--trajectory pairs the VLA generated at the
four preceding steps $t-1,\dots,t-4$ of the same rollout, so each window spans
$0.4$\,s, and run the monitor unchanged on all five pairs. The aggregated alarm $\operatorname{Alarm}_t$ is
scored against the human label of the target; history steps contribute verdicts
only and carry no labels. Besides the single-step rule $\operatorname{Alarm}_t=\hat{y}_t$ of
Table~\ref{tab:main-results}, we evaluate
\begin{align*}
\operatorname{Alarm}^{m/n}_t &= \Bigl[\textstyle\sum_{i=0}^{n-1} \hat{y}_{t-i} \ge m\Bigr],\\
\operatorname{Alarm}^{\mathrm{decay}}_t &= \Bigl[\textstyle\sum_{i=0}^{4} w_i \hat{y}_{t-i} \ge
\tfrac{1}{2}\sum_{i=0}^{4} w_i\Bigr],
\end{align*}
where $[\cdot]$ is the indicator function. We use
$m/n$ of $2/2$, $3/3$, $2/3$, and $3/5$, where $2/2$ and $3/3$ require two and three
consecutive flags, and $w_i=2^{-i/h}$ discounts step $t-i$ with a half-life of
$h=2$ steps ($0.2$\,s). With $h=2$, the decay rule alarms, for example, when both the current and previous steps are 
flagged, and it also alarms with the current step unflagged if the three preceding steps
are flagged. 

We build one window for each of the $100$ pairs in the reliable subset by
taking that pair as the target step and looking back four planning steps. A
window is excluded if the rollout has fewer than four steps before the target,
or if any of its five steps lacks a CoT, a trajectory, or usable lane geometry,
leaving $92$ targets ($25$ \textsc{inconsistent}) shared by all monitors.

\begin{table}[H]
\caption{Temporal aggregation of lane-aware monitor verdicts (judge: Qwen3.5-4B) on the
$92$ targets with complete five-step windows ($25$ \textsc{inconsistent}).
Each rule decides the target step from its verdict and those of up to four
preceding planning steps; Steps is the number of verdicts it reads. The
single-step row is Table~\ref{tab:main-results} restricted to these targets.
Median single-step-F1 run of three repeats with different seeds; best F1 of that run in
\textbf{bold}.}
\label{tab:temporal-agg}
\centering
\small
\setlength{\tabcolsep}{4pt}
\begin{tabular}{@{}lc ccc@{}}
\toprule
& & \multicolumn{3}{c}{Lane-aware (Qwen3.5-4B)} \\
\cmidrule(l){3-5}
Rule & Steps & Prec. & Rec. & F1 \\
\midrule
Single step & 1 & 0.58 & 0.88 & 0.70 \\
\midrule
Two consecutive ($2/2$) & 2 & 0.88 & 0.60 & \textbf{0.71} \\
Three consecutive ($3/3$) & 3 & 0.85 & 0.44 & 0.58 \\
2-out-of-3 vote & 3 & 0.54 & 0.76 & 0.63 \\
3-out-of-5 vote & 5 & 0.47 & 0.68 & 0.56 \\
Half-life decay ($h{=}2$) & 5 & 0.56 & 0.76 & 0.64 \\
\bottomrule
\end{tabular}
\end{table}

Table~\ref{tab:temporal-agg} shows how alarm rules trade precision against
recall. For the lane-aware monitor
(Qwen3.5-4B), requiring two
consecutive flags raises precision from $0.58$ to $0.88$ at a recall cost of
$0.28$ ($0.88\to0.60$), lifting $F_1$ from $0.70$ to $0.71$, while three
consecutive flags recover only $0.44$ of the inconsistencies ($F_1=0.58$).
Voting and decay lower both precision and recall.

These results measure only whether each alarm agrees with the human consistency
label at the target step. Whether acting on these alarms by
triggering a fallback would make driving safer requires monitor-in-the-loop evaluation,
which we leave for future work.

\section{Score Threshold}
\label{app:threshold}
Table~\ref{tab:threshold-sweep} varies $\theta$ from $1$ to $4$ on the same runs as
Table~\ref{tab:main-results}. Because these runs are scored against the evaluation
labels, the sweep shows sensitivity to $\theta$.
Section~\ref{sec:discussion} discusses the resulting trade-off.

\begin{table*}[t]
\caption{Effect of the score threshold on judge-based detection. A pair is
flagged \textsc{inconsistent} when the judge's 1--5 consistency score is at most
the threshold; Table~\ref{tab:main-results} uses threshold~2. Each cell reports
precision / recall / F1 with \textsc{inconsistent} as the positive class.
Each row shows the representative median-F1 run of three repeats, the same
runs as Table~\ref{tab:main-results}.}
\label{tab:threshold-sweep}
\centering
\small
\setlength{\tabcolsep}{5pt}
\begin{tabular}{@{}ll cccc@{}}
\toprule
Judge & Monitor & Threshold 1 & Threshold 2 & Threshold 3 & Threshold 4 \\
\midrule
Qwen3.5-4B & Ego-frame & 0.45 / 0.35 / 0.39 & 0.44 / 0.85 / 0.58 & 0.43 / 0.88 / 0.58 & 0.33 / 0.92 / 0.48 \\
           & Lane-aware & 0.85 / 0.42 / 0.56 & 0.59 / 0.88 / 0.71 & 0.55 / 0.88 / 0.68 & 0.45 / 0.96 / 0.62 \\
\midrule
DeepSeek V4.1 Flash & Ego-frame & 0.50 / 0.31 / 0.38 & 0.54 / 0.73 / 0.62 & 0.50 / 0.77 / 0.61 & 0.41 / 0.88 / 0.56 \\
              & Lane-aware & 0.79 / 0.42 / 0.55 & 0.75 / 0.81 / 0.78 & 0.68 / 0.88 / 0.77 & 0.44 / 0.92 / 0.59 \\
\midrule
Opus 5.5      & Ego-frame & 0.54 / 0.27 / 0.36 & 0.55 / 0.69 / 0.61 & 0.48 / 0.85 / 0.61 & 0.35 / 1.00 / 0.52 \\
              & Lane-aware & 0.88 / 0.27 / 0.41 & 0.84 / 0.81 / 0.82 & 0.60 / 0.96 / 0.74 & 0.34 / 1.00 / 0.50 \\
\midrule
Jev 1.14      & Ego-frame & 0.41 / 0.35 / 0.38 & 0.43 / 0.81 / 0.56 & 0.40 / 0.85 / 0.54 & 0.32 / 0.88 / 0.46 \\
              & Lane-aware & 0.77 / 0.38 / 0.51 & 0.51 / 0.88 / 0.65 & 0.47 / 0.96 / 0.63 & 0.30 / 0.96 / 0.46 \\
\bottomrule
\end{tabular}
\end{table*}

\end{document}